%% file: sp2025.tex
\documentclass[conference,compsoc]{IEEEtran}

\input{preamble.tex}
\input{commands.tex}

\begin{document}
\IEEEoverridecommandlockouts

\title{Comet: Accelerating Private Inference for Large Language Model by Predicting Activation Sparsity}


\author{\IEEEauthorblockN{
Guang Yan\IEEEauthorrefmark{2}, 
Yuhui Zhang\IEEEauthorrefmark{2}\IEEEauthorrefmark{1}, 
Zimu Guo\IEEEauthorrefmark{2}, 
Lutan Zhao\IEEEauthorrefmark{2}, 
Xiaojun Chen\IEEEauthorrefmark{2}, 
Chen Wang\IEEEauthorrefmark{3}, 
Wenhao Wang\IEEEauthorrefmark{2},  \\
Dan Meng\IEEEauthorrefmark{2} and 
Rui Hou\thanks{\IEEEauthorrefmark{1} Yuhui Zhang and Rui Hou are the corresponding authors.}\IEEEauthorrefmark{2}\IEEEauthorrefmark{1}}
    \IEEEauthorblockA{\IEEEauthorrefmark{2}State Key Laboratory of Cyberspace Security Defense, Institute of Information Engineering, CAS\\ 
    and University of Chinese Academy of Sciences\\
    \{yanguang1997, zhangyuhui, guozimu, zhaolutan, chenxiaojun, wangwenhao, mengdan, hourui\}@iie.ac.cn \\  
    \IEEEauthorrefmark{3}EIRI, NELBDRC, Tsinghua University, wang\_chen@tsinghua.edu.cn }
}

\maketitle
\pagestyle{empty} 

\begin{abstract}
    With the growing use of large language models (LLMs) hosted on cloud platforms to offer inference services, privacy concerns about the potential leakage of sensitive information are escalating. 
    Secure Multi-Party Computation (MPC) is a promising solution to protect the privacy in LLM inference. However, MPC requires frequent inter-server communication, causing high performance overhead.

    Inspired by the prevalent activation sparsity of LLMs, where most neuron are not activated after non-linear activation functions, we propose an efficient private inference system, {\projectname}. This system employs an accurate and fast predictor to predict the sparsity distribution of activation function output. Additionally, we introduce a new private inference protocol. It efficiently and securely avoids computations involving zero values by exploiting the spatial locality of the predicted sparsity distribution. 
    While this computation-avoidance approach impacts the spatiotemporal continuity of KV cache entries, we address this challenge with a low-communication overhead cache refilling strategy that merges miss requests and incorporates a prefetching mechanism.
    Finally, we evaluate {\projectname} on four common LLMs and compare it with six state-of-the-art private inference systems. {\projectname} achieves a $1.87\times$-$2.63\times$ speedup and a $1.94 \times$-$2.64\times$ communication reduction.
\end{abstract}

\input{chapters/1_introduction}
\input{chapters/2_background}
\input{chapters/3_system}

\input{chapters/4_predictor}
\input{chapters/5_spmm}
\input{chapters/6_kv_cache}
\input{chapters/8_evaluation}
\input{chapters/9_related_work}

\input{chapters/11_conclusion}

\bibliographystyle{IEEEtranS}
\bibliography{references}

\appendices

\input{chapters/7_security}

\input{chapters/appendix_optimality}

\input{chapters/appendix_generality}

\input{chapters/appendix_baselines}
\input{chapters/10_discussion}
\input{chapters/meta_review}

\end{document}

%% file: preamble.tex
\usepackage[nocompress]{cite}
\usepackage{textcomp}
\usepackage{xcolor}
\usepackage[hyphens]{url}
\usepackage{fancyhdr}

\usepackage{CJK}

\usepackage{amsmath, amsthm, amsfonts, amssymb}
\usepackage{bm}

\usepackage{graphicx}
\usepackage{multirow}
\usepackage{caption, subcaption}
\usepackage{tabularx}
\usepackage{booktabs}

\usepackage{enumerate}
\usepackage{enumitem}
\setlist[enumerate,1]{leftmargin=3em}

\usepackage{pifont}

\usepackage{color}

\usepackage{algorithm}
\usepackage{algorithmicx}
\usepackage{algpseudocode}
\floatname{algorithm}{Protocol}

\usepackage{float}

\usepackage{hyperref}
\hypersetup{
    colorlinks = true,
    citecolor = black, 
    linkcolor=black,
}

\theoremstyle{plain}
\newtheorem{theorem}{Theorem}

\theoremstyle{definition}

\theoremstyle{remark}

\usepackage{placeins}
\usepackage{xspace}

\newcolumntype{P}[1]{>{\centering\arraybackslash}p{#1}}
\newcolumntype{C}{>{\color{blue}}c}

\usepackage{ulem}

%% file: commands.tex
\algnewcommand\algorithmicinput{\textbf{Input:}}
\algnewcommand\Input{\item[\algorithmicinput]}
\algnewcommand\algorithmicoutput{\textbf{Output:}}
\algnewcommand\Output{\item[\algorithmicoutput]}

\algnewcommand\algorithmicpreprocess{\textbf{Preprocess:}}
\algnewcommand\Preprocess{\item[\algorithmicpreprocess]}

\algnewcommand\algorithmicglobal{\textbf{Global Parameters:}}
\algnewcommand\Global{\item[\algorithmicglobal]}

\algnewcommand{\Continue}{\State \textbf{continue}}
\algnewcommand{\Break}{\State \textbf{break}}

\newcommand{\projectname}{\emph{Comet}}

\newcommand\bx{\mathbf{x}}
\newcommand\by{\mathbf{y}}

\newcommand\ba{\mathbf{a}}
\newcommand\bb{\mathbf{b}}
\newcommand\bc{\mathbf{c}}

\newcommand{\bX}{\mathbf{X}}
\newcommand{\bY}{\mathbf{Y}}
\newcommand{\bW}{\mathbf{W}}
\newcommand{\bZ}{\mathbf{Z}}

\newcommand{\bE}{\mathbf{E}}
\newcommand{\bF}{\mathbf{F}}

\newcommand{\share}[1]{\ensuremath{[\![ #1 ]\!]}\xspace}
\newcommand{\shareB}[1]{\ensuremath{[\![ \mathbf{#1} ]\!]}\xspace}

\newcommand{\mypara}[1]{\vspace{2pt}\noindent\textbf{{#1}}}

%% file: chapters/1_introduction.tex
\section{Introduction}

Large Language Models (LLMs) have been extensively utilized in a range of tasks such as text generation, question answering, sentiment analysis and reading comprehension~\cite{brown2020language, touvron2023llama, yuan2023distilling, bommasani2021opportunities, liang2022holistic, chan2022data}. 
Currently, LLMs commonly follow the ``Deep Learning as a Service'' (DLaaS) paradigm~\cite{soifer2019deep}, wherein LLMs are deployed on cloud servers as service providers, and users send input data to these servers to perform inference tasks. 
However, this raises privacy concerns, as the user's data may be privacy-sensitive.

One promising solution to address the privacy concerns in LLM inference is Secure Multi-Party Computation (MPC), known as MPC-based private inference~\cite{dong2023puma, mohassel2018aby3, knott2021crypten, kumar2020cryptflow, wagh2019securenn, wagh2020falcon, tan2021cryptgpu, riazi2018chameleon, patra2020blaze, byali2019flash, chaudhari2019trident, maeng2024accelerating, li2022mpcformer}. 
Specifically, the model owner and the user provide their models or inputs to multiple non-colluding ($\ge 2$) MPC servers in a secret-shared form, then servers execute the private inference protocol and send the results back to the user.
This method ensures that no single server can recover the original data, thereby enabling privacy-preserving inference.

However, private inference encounters significant performance problem, primarily due to extensive communication between MPC servers.
These servers frequently exchange intermediate results derived from their local computations. 
For instance, in the case of OPT-6.7B~\cite{zhang2022opt}, communication time constitutes over 85\% of the total inference time. 
Prior works have mainly focused on reducing the communication overhead of non-linear layers by designing model architectures with fewer non-linear operations~\cite{li2022mpcformer, akimoto2023privformer, liu2023llms}, or by using more efficient approximations for non-linear functions~\cite{dong2023puma, luo2024secformer, chen2024securetlm}. The optimization for linear layers has been overlooked, especially in private LLM inference, where this part of communication incurs higher overhead.

\begin{figure}[htbp]
    \centering
    \includegraphics[width=0.42\textwidth]{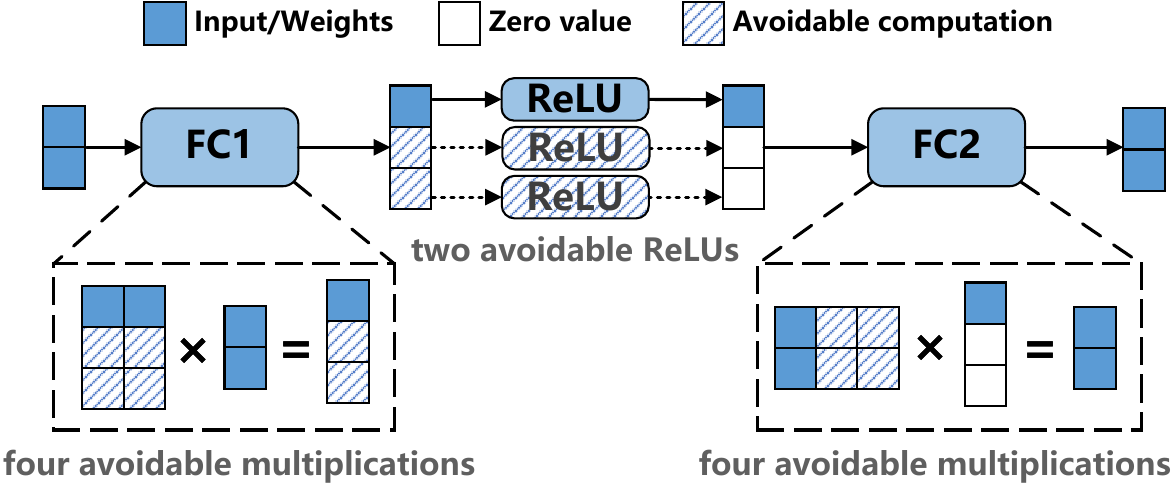}
    \caption{Activation sparsity of LLMs.}
    \label{fig:motivation}
    \vspace{-0.5cm}
\end{figure}

Fortunately, the activation sparsity~\cite{liu2023deja, song2023powerinfer, song2024prosparse, zhang2024relu, mirzadeh2023relu} of LLMs provides an opportunity to accelerate both linear and non-linear computations. For instance, in OPT-6.7B, more than 90\% of the neurons output zero after the activation function (ReLU) of its Feed-Forward Network (FFN). 
As illustrated in Figure~\ref{fig:motivation}, if activation sparsity can be accurately predicted (depicted by white boxes), it becomes possible to identify neurons that will output zero in advance. Consequently, the computations of their non-linear activation functions can be omitted, and the related linear computations in both preceding and subsequent linear layers can also be skipped. By avoiding these unnecessary computations (shaded boxes), the associated communication overhead in private inference can also be reduced.

\noindent\textbf{Contributions.} This paper proposes {\projectname}, an efficient private inference system that leverages activation sparsity to reduce computation and communication overhead in MPC. Unlike prior work focusing on optimizing specific MPC protocols, {\projectname} is orthogonal to these efforts and integrates the novel prediction mechanism and the computaion-communication avoidance mechanism into the classical private inference framework. To the best of our knowledge, this is the first system to exploit activation sparsity for accelerating private inference in LLMs. The main contributions of this paper are as follows:

\begin{itemize}[leftmargin=1em]
    \item We propose a predictor to estimate the sparsity distribution of activation outputs. To ensure accurate prediction and efficient execution, we implement the predictor using a lightweight neural network. To preserve privacy, the predictor generates a secret-shared sparsity distribution, which is collaboratively computed by MPC servers using secure protocols. However, indexing activated neurons based on this secret-shared distribution poses efficiency challenges, as existing secure indexing methods require numerous expensive comparison operations~\cite{li2021privacy}, negating the efficiency gains of sparsity. To address this, we introduce a shuffle-based indexing technique that uses oblivious shuffling to randomize the order of the sparsity distribution, allowing plaintext indexing while preserving privacy. 
    These designs significantly reduce prediction and indexing latency, establishing the feasibility of sparsity prediction within private inference systems.

    \item We propose a private inference protocol designed to minimize computation and communication overhead by leveraging the spatial locality of the sparsity distribution.  
    For the linear layers preceding and following the activation function, computation can be reduced by identifying nonzero elements of the sparsity distribution and performing only the corresponding dot products.
    However, in classical private inference protocols, naively performing dot products separately results in redundant communication due to the need for repeated masking and communicating the rows or columns of the matrix. 
    In contrast, our protocol fully exploits the spatial locality of sparsity distribution by grouping the nonzero elements resulting from dot products of the same matrix rows or columns into a single block, ensuring that each row or column is masked and communicated only once. This approach enhances the efficiency of secure computations, achieving up to 300$\times$ reduction in communication overhead compared to classical methods.

    \item We propose a KV cache refilling strategy to address the compatibility challenges between sparsity prediction and KV caching. KV cache~\cite{li2024survey} is a core optimization technique in LLM systems, enhancing performance by storing and reusing intermediate results. However, the computation savings introduced by sparsity prediction disrupt the spatiotemporal continuity of cached key-value pairs, reducing the cache hit rate and requiring costly refills. To mitigate this, our strategy merges consecutive cache miss requests, often targeting the same attention heads or tokens, and incorporates a prefetching mechanism to proactively handle potential misses. 
    These designs improve the compatibility between sparsity-aware computation and KV caching, enabling efficient integration with modern private LLM inference systems.
\end{itemize}

Finally, we evaluate the performance of {\projectname} system using four different sizes of LLMs. The results indicate that, compared to six state-of-the-art private inference systems, the {\projectname} achieves $1.87\times$-$2.63\times$ end-to-end speedup and $1.94\times$-$2.64\times$ communication reduction.

%% file: chapters/2_background.tex
\section{Understanding of Private Inference} 

\subsection{MPC-based private inference}

Secure Multi-Party Computation (MPC)~\cite{cramer2015secure} is a cryptographic technique that provides a promising solution for private inference among multiple participants~\cite{li2022mpcformer,dong2023puma,gupta2023sigma,luo2024secformer,knott2021crypten, riazi2018chameleon, kumar2020cryptflow}. It relies on cryptographic primitives such as secret sharing to protect both model weights and inference data~\cite{shamir1979share, blakley1979safeguarding, goldreich2019play}. This work adopts an additive secret sharing scheme, which is widely used in private inference due to its efficiency and applicability.

\mypara{Additive secret sharing}. For simplicity, we illustrate additive secret sharing in a two party setting ($P_1$ and $P_2$):

\begin{itemize}[leftmargin=1em]
    \item \textit{Sharing}:  A secret $x$ is split into two shares $\share{x}_1$ and $\share{x}_2$ such that $x = \share{x}_1 + \share{x}_2$, ensuring no single share can retrieve information of the secret.  The data owner sends $\share{x}_1$ to $P_1$ and $\share{x}_2$ to $P_2$.
    \item \textit{Reconstruction}: To reveal the secret $x$, $P_1$ and $P_2$ exchange their shares and locally compute $x = \share{x}_1 + \share{x}_2$.
    \item \textit{Addition}:  To compute $z=ax+by+c$, where $a,b,c$ are public values and $x,y$ are secret-shared among $P_1$ and $P_2$, each party locally computes $\share{z}_i = a\share{x}_i + b\share{y}_i + (i-1)c$.
    \item \textit{Multiplication}:  To compute $z=xy$ where $x,y$ are secret shared among $P_1$ and $P_2$, the parties use pre-generated Beaver triples $(\share{a}, \share{b}, \share{c})$ where $c = a \cdot b$. The triples can be provided by a trusted third party or generated via cryptographic methods such as homomorphic encryption~\cite{paillier1999public} or oblivious transfer~\cite{keller2016mascot}. Each party locally computes differences $\share{d}_i = \share{x}_i - \share{a}_i$ and $\share{e}_i = \share{y}_i - \share{b}_i$, reconstructs $d$ and $e$, and computes: 
        \begin{equation}
            \label{eq:multiplication}
            \share{z}_i = \share{c}_i + d \cdot \share{b}_i + e \cdot \share{a}_i + (i-1) \cdot d \cdot e
        \end{equation}
\end{itemize}
Based on the above basic operations, various protocols can be constructed, including dot product ($\Pi_\textsf{DP}$) and matrix multiplication ($\Pi_\textsf{MatMul}$). In this paper, we employ them in a blackbox manner.

\mypara{Private inference.} In a MPC-based private inference system, three primary roles are typically identified: the model owner, the user, and at least two non-colluding MPC servers. 
In practical, the computing nodes from different cloud platforms are usually selected to act as MPC servers to meet the requirement of non-colluding~\cite{apple2021exposure, mpc_alliance}. Additionally, if the user has enough compute capability, it can serve as one of the MPC servers~\cite{dong2023puma, maeng2024accelerating}.
This setup ensures that at least one server remains independent, thereby further enhancing the guarantee of non-colluding.

Before inference,  the model owner uses secret sharing to splits the model parameters into shares, deploying them across MPC servers. During inference, the user also splits the input and distributes shares to the servers.
MPC servers execute each layer of models sequentially using the input share and model share. Specifically, they jointly complete each basic operation within a layer, such as addition, multiplication and comparison through MPC protocols.
The result of each basic operation is still distributed in the form of shares on MPC servers and is used for subsequent computations. After the completion of the inference, the result shares are returned to the user for reconstructing the final result.

\subsection{LLM private inference has high communication overhead}

\begin{figure}[htbp]
    \vspace{-0.2cm}
    \centering
    \includegraphics{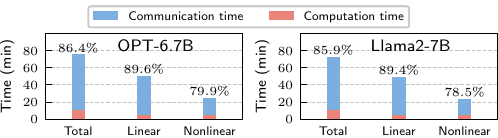}
    \caption{MPC-based private inference time breakdown with an input length of 512, output length of 16, and bandwidth of 5Gbps. The numbers on the bars is the proportion of communication time.}
    \label{fig:inference_overhead}
    \vspace{-0.2cm}
\end{figure}

Although MPC-based private inference can protect the privacy of user data and models, it faces critical challenges when applied to LLMs. A key limitation of MPC protocols is their reliance on frequent exchanges of intermediate results between servers, which makes communication overhead the primary performance bottleneck. The deep architectures and massive parameter sizes of LLMs further exacerbate this issue: linear layers, containing billions of parameters, require extensive communication, while non-linear layers incur significant overhead as their non-linear functions often need to be approximated by high-degree polynomials or implemented using complex logic operations~\cite{kumar2020cryptflow,chen2024securetlm, wagh2020falcon, knott2021crypten}, both of which are costly for high-dimensional activations.
As illustrated in Figure~\ref{fig:inference_overhead}, both OPT-6.7B~\cite{zhang2022opt} and Llama2-7B~\cite{song2024prosparse} require over 60 minutes to generate 16 tokens, with communication time comprises over 90\% of the total inference time. Moreover, for both non-linear and linear layers, communication accounts for the majority of the overhead, exceeding 80\%.

\subsection{Prevalent activation sparsity enables significant reductions in communication for private inference}

LLMs typically consist of multiple Transformer blocks~\cite{vaswani2017attention}. Each Transformer block includes a Multi-Head Attention (MHA) module and a Feed-Forward Network (FFN), both of which contain two linear layers with an intermediate non-linear layer. The non-linear layer in the MHA is the Softmax function applied in the attention heads, while the FFN uses the ReLU~\cite{nair2010rectified} function.

\begin{figure}[h]
    \centering  \includegraphics{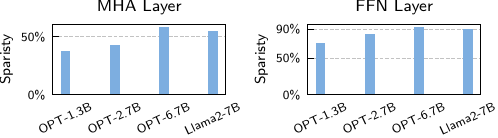}
    \caption{Activation sparsity of different LLMs, evaluated on the same dataset used in Section~\ref{sec:evalution}.}
    \label{fig:motivation_sparsity}
\end{figure}

A key property of LLMs is their \textbf{activation sparsity}—a phenomenon where many outputs of non-linear layers are zero or near zero~\cite{liu2023deja, song2023powerinfer, song2024prosparse, zhang2024relu, mirzadeh2023relu}. For instance, many attention heads in the MHA may remain inactive (i.e., the norm of their output vectors is close to zero), and the ReLU function in the FFN produces many zeros. Even for LLMs that use activation functions without inherent sparsity, such as GeLU~\cite{hendrycks2016gaussian} and SwiGLU~\cite{shazeer2020glu}, activation sparsity can still be achieved by replacing the activation function with ReLU and fine-tuning, with a slight accuracy loss~\cite{zhang2024relu, song2024prosparse, song2024turbo}. 
Our analysis of various LLMs highlights the prevalence of activation sparsity, particularly in larger models. As shown in Figure~\ref{fig:motivation_sparsity}, for OPT and Llama2, more than 50\% of attention heads in the MHA of are inactive, while over 90\% of ReLU outputs in the FFN are zero. We provide more results on activation sparsity across different model architectures and different activation functions in Appendix~\ref{sec:generality}.

\begin{figure}[h]
    \centering
    \includegraphics{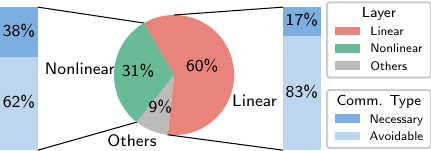}
    \caption{Communication cost breakdown for OPT-6.7B with an input length of 512 and output length of 16.}
    \label{fig:motivation_breakdown}
\end{figure}

This property allows us to disregard non-linear computations without affecting the final output. Furthermore, this optimization cascades: when a non-linear computation can be skipped, the computations in the preceding linear layer that generate its input, as well as those in the subsequent linear layer that consume its result, can also be avoided.
This reduction in computation directly translates to a reduction in communication overhead for private inference. Profiling the communication of OPT-6.7B during a single private inference reveals that 31\% of communication traffic comes from non-linear layers, with 62\% of it avoidable; 60\% comes from linear layers, with 83\% avoidable. Overall, over 70\% of the communication traffic for private inference can be eliminated, as shown in Figure~\ref{fig:motivation_breakdown}.

%% file: chapters/3_system.tex
\begin{figure*}
    \centering
    \includegraphics[width=0.95\textwidth]{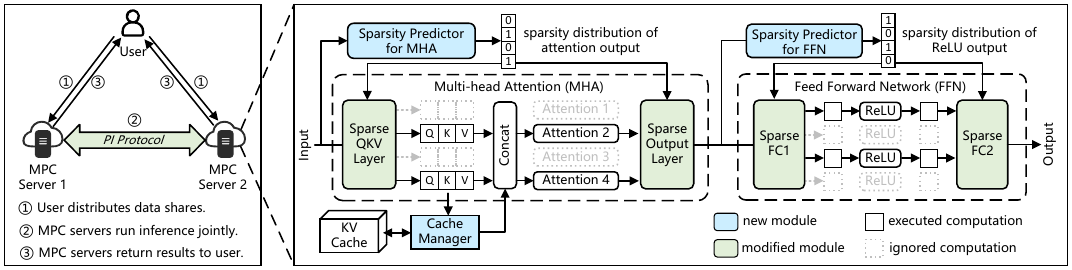}
    \caption{System overview of {\projectname}.}
    \label{fig:system_overview}
\end{figure*}  

\section{{\projectname} System Overview} 

\subsection{Design motivation} 

\begin{figure}[h]
    \centering
    \includegraphics[width=0.47\textwidth]{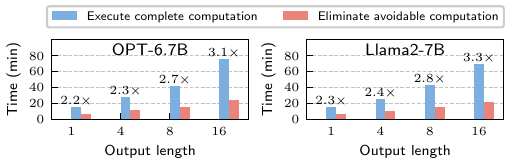}
    \vspace{-0.2cm}
    \caption{Ideal speedup.}
    \label{fig:motivation_speedup}
    \vspace{-0.5cm}
\end{figure}

As previously noted, LLMs commonly exhibit activation sparsity, where many attention heads in MHA are not activated and numerous FFN ReLU outputs are zero. By accurately predicting this sparsity, we can omit the corresponding non-linear computations. Additionally, computations in the preceding linear layers generating these zero values, as well as those in the subsequent linear layers processing them, can also be skipped.
As shown in Figure~\ref{fig:motivation_speedup}, eliminating these computations under ideal conditions (i.e., perfect prediction) can accelerate inference for models such as OPT-6.7B and Llama2-7B by approximately $2.2\times$-$3.3\times$. This speedup becomes even more significant with longer generated sequences.
Inspired by this observation, we propose {\projectname}, a private inference system designed to leverage activation sparsity prediction to minimize unnecessary computations and reduce communication overhead.

\subsection{Threat model}
{\projectname} adopts a standard MPC threat model, where predefined programs run among multiple parties, protecting input data and intermediate results while typically revealing only the final outputs to designated parties. All correlated randomness required for the protocol execution is generated and distributed by a trusted third party (TTP).
Following prior works~\cite{li2022mpcformer,gupta2023sigma,luo2024secformer, knott2021crypten, riazi2018chameleon, kumar2020cryptflow}, we assume an honest-but-curious adversary who passively corrupts one or more MPC servers but cannot corrupt all servers. Such an adversary strictly follows the system protocols but may attempt to infer sensitive information from the data it observes.

To leverage activation sparsity, we assume that sparsity level (i.e., the number of zero values) can be revealed, while the sparsity distribution (i.e., the positions of zero values) remains hidden. This assumption is practical in real-world applications of LLM inference because activation sparsity levels are primarily determined by model size—a publicly disclosed parameter by major model providers such as Meta~\cite{touvron2023llama} and Google~\cite{team2024gemma}—and exhibit small variation across different inputs~\cite{li2022lazy, song2024prosparse}. This assumption also aligns with other works that exploit sparsity in MPC~\cite{schoppmann2019make, cui2021exploiting, bitar2024sparsity}. 

\subsection{Design overview}  

{\projectname} aims to provide efficient and privacy-preserving inference for LLMs by exploiting activation sparsity. The system accelerates computation and reduces communication while ensuring input and model privacy through MPC. Figure~\ref{fig:system_overview} illustrates an overview of {\projectname}. It comprises a user and multiple MPC servers holding secret shares of the LLM. Before inference, the user distributes secret shares of the input to each MPC server. These servers then collaboratively execute the inference task using private inference protocols. Finally, the servers return the result shares to the user, who reconstructs the final inference result.

To predict activation sparsity, {\projectname} employs neural network-based \emph{predictors} for both the MHA and FFN. The predictors are designed to be shallow and low-rank, enabling rapid computation during inference, and are pre-trained locally on large public datasets to ensure accuracy. 
The predictors are secret-shared and deployed across multiple MPC servers. The servers collaboratively execute these predictors via MPC protocols to obtain secret shares of the sparsity distribution.
To protect the sparsity distribution, {\projectname} employs a cryptographic technique, oblivious shuffle, to obscure the positions of nonzero elements. {Additionally, a secure indexing mechanism is proposed to efficiently retrieve elements from the shuffled sparsity distribution.}

With the predicted sparsity distribution, {\projectname} identifies non-activated attention heads in MHA and zero ReLU outputs in FFN to skip their associated computations and communications. To achieve this, the private inference protocol is extended to support \emph{sparse matrix multiplication} for both the preceding linear layers (e.g., Sparse QKV Layer and Sparse FC1) and the subsequent linear layers (e.g., Sparse Output Layer and Sparse FC2) around the nonlinear layers. For the nonlinear layers themselves, which involve element-wise or vector-wise operations, the protocol remains unchanged, as computations can be directly skipped by filtering inputs corresponding to zero values.

The sparsity-aware computation mechanism in the Sparse QKV Layer omits certain key and value computations, leading to missing KV entries. When these entries are needed in subsequent inference, their absence can interrupt the process or, if ignored, significantly degrade accuracy. To address this compatibility challenge between sparsity prediction and KV cache, {\projectname} introduces a KV \emph{cache manager} that efficiently manages missing entries through cache refilling and prefetching strategies, reducing the overhead of refilling while ensuring seamless operation of the KV caching mechanism.

\mypara{Workflow}. We take MHA as an example to illustrate the workflow of {\projectname}, as FFN operates similarly.
First, the MHA \emph{predictor} determines activated attention heads. Then, the QKV layer computes for these heads using \emph{sparse matrix multiplication} and outputs their query, key, and value. The \emph{cache manager} stores keys and values in the KV cache and checks for misses. If found, cache miss requests are merged and the cache is refilled. Afterward, attention computations are performed only for the activated heads, generating the attention output. This is followed by \emph{sparse matrix multiplications} in the output linear layers, where computations corresponding to non-activated heads are skipped. Finally, the output are reshaped to origin size for the subsequent layer computation.

%% file: chapters/4_predictor.tex
\section{Activation Sparsity Predictor with Oblivious Shuffle}
\label{subsec:sparsity_predictor}

\subsection{Basic design}

To predict activation sparsity efficiently, {\projectname} incorporates two predictors per Transformer layer, one for the MHA module and one for the FFN. The predictors use the inputs to the MHA or FFN as their inputs, as these inputs fully determine the activation states during inference. This is because model parameters remain fixed during inference, making the input data the primary factor influencing activation states.

\begin{figure}[htbp]
    \centering
    \includegraphics{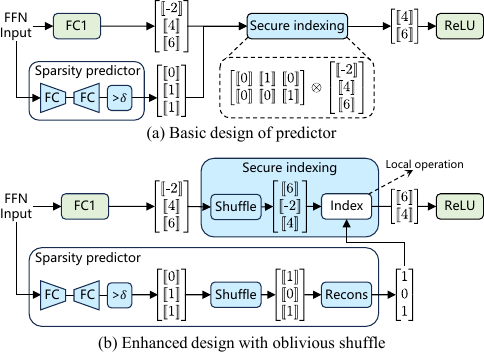}
    \caption{Activation sparsity predictor and secure indexing mechanism for FFN. The design for MHA is similar.}
    \label{fig:predictor_arch}
    \vspace{-0.4cm}
\end{figure}

Both predictors share the same architecture. Taking the FFN as an example, as illustrated in Figure~\ref{fig:predictor_arch}(a), the predictor is a shallow neural network consisting of two fully connected layers (FC) with low-rank weight matrices, followed by a threshold layer. It can be formulated as $\by = \sigma(\bW_2( \bW_1 \bx + \bb_1 ) + \bb_2)$, where $\bx$ is the FFN input and $\sigma$ is a threshold function that outputs 1 if $\sigma(x) > \delta$, otherwise 0. The low-rank structure is chosen because activation outputs often reside in a low-dimensional subspace, making it sufficient to capture sparsity patterns. The threshold layer compares the outputs of the previous layer against a predefined threshold $\delta$ to determine activation. For the FFN, it determines whether each neuron outputs a nonzero value after the activation function. For MHA, it determines whether the norm of each attention head's output vector is significantly above zero.

The predictors are pre-trained by the model owner using publicly available datasets~\cite{merity2016pointer, raffel2020exploring}. Training data is generated by collecting the inputs of MHA and FFN and outputs from the activation functions of MHA and FFN during inference.

\subsection{Enhancing indexing efficiency with oblivious shuffle}

By leveraging the sparsity distribution to index inputs that lead to nonzero outputs in non-linear layers, computations can be focused on these inputs, reducing overall computation and communication overhead.

A naive indexing approach is to reveal the secret-shared sparsity distribution as plaintext and then index inputs directly.
However, this approach compromises privacy. If the sparsity distribution is exposed, it could leak sensitive patterns in the data, opening the system to privacy risks such as inference attacks~\cite{carlini2022membership}. For example, by sending known inputs and observing their sparsity patterns, an attacker could compare these patterns to a ciphertext input to determine whether it is similar to the known inputs. Additionally, attackers could analyze the sparsity distribution of a ciphertext input to infer whether it belongs to a specific dataset.

To address the privacy concerns, secure indexing methods have been proposed~\cite{li2021privacy, schoppmann2019make}. 
However, existing methods 
face significant efficiency challenges when applied to latency-sensitive private inference.
For OPT-6.7B, secure indexing the ReLU inputs requires roughly 1000$\times$ the execution time of the ReLU itself, outweighing any potential time savings gained through sparsity. Existing methods typically consist of three steps: (1) Assume that the indexing vector $\shareB{s} \in \mathbb{R}^n$ contains $m$ nonzero elements. The argmax protocol is executed $m$ times to obtain $m$ secret indices. (2) For each secret index $\share{i}$, a secret one-hot vector $\shareB{i}$ is created by executing the equality protocol $n$ times. This vector has $\share{1}$ at the position $\share{i}$ and $\share{0}$ elsewhere. All one-hot vectors are then concatenated into a matrix $\shareB{I} \in \mathbb{R}^{m \times n}$. (3) Finally, a secure multiplication is performed between $\shareB{I}$ and the target vector $\shareB{x} \in \mathbb{R}^n$ to obtain the result vector $\shareB{o} \in \mathbb{R}^{m}$. The total communication cost is $(m\log n + 2mn)C + 2mn$, where $C$ is the communication cost of a comparison operation.

To address the inefficiencies of secure indexing while maintaining privacy, we propose a novel shuffle-based indexing technique that allows plaintext indexing without compromising security. The key insight is that the privacy risks of plaintext indexing arise from exposing the positions of nonzero elements in the sparsity distribution. To mitigate this, we incorporate a cryptographic technique, oblivious shuffle~\cite{attrapadung2021oblivious, chase2020secret, laur2011round, movahedi2015secure, chida2019efficient}, to randomize the order of the secret-shared sparsity distribution. This shuffled distribution can then be revealed in plaintext without exposing the original sparsity pattern. As illustrated in Figure~\ref{fig:predictor_arch}(b), the sparsity distribution output by the threshold layer is shuffled and then reconstructed as plaintext. To ensure consistent indexing, the target vector (FC1 output) is also shuffled to match the revealed distribution's order, allowing indexing to be performed locally. 
While the sparsity pattern remains hidden, the sparsity levels are still revealed. Although this information does not expose specific inputs, repeated observations could potentially leak privacy. This can be mitigated by adding noise to the sparsity levels using MPC-based differential privacy \cite{NIPS2015_a0161022}. A more detailed discussion is provided in Appendix~\ref{appendix:dp}.

\mypara{Oblivious shuffle.} The shuffle operation in the predictor is implemented using Protocol~\ref{protocol:shuffle}, which takes only one communication round at the online stage, with a communication cost equal
to the size of the shuffled data itself. Therefore, the communication cost of our shuffle-based indexing method is only $2n$. Compared to traditional secure indexing methods, our method significantly reduces both computation and communication overhead while preserving the privacy of sparsity patterns of activation during inference.

Oblivious shuffle protocol reorders secret-shared vector $\shareB{x}$ according to a secret random permutation $\pi$. For the permutation $\pi = [2,1,3]$, $\pi(\mathbf{x})= [\mathbf{x}_2, \mathbf{x}_1, \mathbf{x}_3]$.
The core idea of oblivious shuffle is the decomposition of $\pi$ into two interdependent sub-permutation pairs $(\rho_i, \tau_i), i=1,2$, satisfying that $\pi = \rho_1 \circ \tau_2 = \rho_2 \circ \tau_1$. Each party $P_i$ only holds $(\rho_i, \tau_i)$, ensuring that neither can reconstruct $\pi$ independently. By first applying $\tau_i$ to their shares, exchanging the masked results, and then applying $\rho_i$ to the received results while removing the masks, the parties collaboratively achieve the effect of applying $\pi$ without revealing $\pi$ or their respective shares. The security proof is in Appendix~\ref{sec:security_analysis}.

 In particular, (i) the sub-permutation $(\rho_i, \tau_i)$ can be reused, requiring only new masks for protection each time. Moreover, applying oblivious shuffle with the same sub-permutation on two secret-shared vectors aligns them in a same and secret order.
(ii) A secret share shuffled by $(\rho_i, \tau_i)$ can be restored to its original order, by each party locally generating the inverse permutations $(\tau_i^{-1}, \rho_i^{-1})$ and performing an oblivious shuffle based on these inverses. We refer to this operation as ``unshuffle".

\begin{algorithm}[h]
    \caption{Oblivious Shuffle Protocol $\Pi_{\mathsf{Shuffle}}$}\label{protocol:shuffle}
    \begin{algorithmic}[1]
        \Input
        $P_i$ holds the share $\share{\mathbf{x}}_i$, the permutation $(\rho_i, \tau_i)$, and the masks $(\ba_i, \bb_i)$, $i\in \{1,2\}$.
        \Output
        $P_i$ gets the share $\share{\mathbf{z}}_i$, where $\mathbf{z}=\pi(\mathbf{x})$, $\pi= \rho_1 \circ \tau_2 = \rho_2 \circ \tau_1$.

        \State \uline{\textbf{At the offline stage:}}
        \State TTP generates random permutation $\pi$, $\tau_1$, $\tau_2$.
        \State TTP computes $\rho_1 = \pi \circ \tau_2^{-1}$, $\rho_2 = \pi \circ \tau_1^{-1}$.
        \State TTP generates random vectors $\ba_1$, $\ba_2$, $\bc$.
        \State TTP computes $\bb_1 = \rho_1(\ba_2) + \bc$, $\bb_2 = \rho_2(\ba_1) - \bc$.
        \State TTP sends $(\rho_i, \tau_i, \ba_i, \bb_i)$ to $P_i$.
        \State \uline{\textbf{At the online stage:}}
        \State $P_i$ computes $\share{\mathbf{y}}_i = \tau_i(\share{\mathbf{x}}_i) + \ba_i$.
        \State $P_i$ sends $\share{\mathbf{y}}_i$ to $P_{i+1}$.
        \State $P_i$ computes $\share{\mathbf{z}}_i = \rho_i(\share{\mathbf{y}}_{i+1}) - \bb_i$.
        \State \Return $\share{\mathbf{z}}_i$
    \end{algorithmic}
\end{algorithm}
\vspace{-0.3cm}

%% file: chapters/5_spmm.tex
\section{Private Inference Protocol Exploiting Spatial Locality of Sparsity Distribution}
\label{sec:sparse_inference}

\begin{figure}[htbp]
    \centering
    \includegraphics[width=0.45\textwidth]{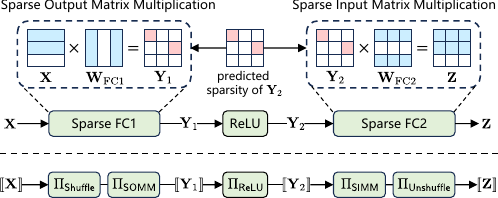}
    \caption{Layers and protocols design for FFN in {\projectname}. The design for MHA is similar.}
    \label{fig:SPMM}
    \vspace{-0.4cm}
\end{figure}

{\projectname} leverages the predicted activation sparsity to accelerate inference by eliminating zero-related computations and communications. For non-linear layers, computations are performed only on inputs with nonzero outputs. For linear layers, only dot products involving nonzero outputs or inputs are computed, as shown in Figure~\ref{fig:SPMM}.

However, in MPC, redundant communication occurs when the same elements of a matrix are involved in multiple separate dot products. This happens because the same elements must be repeatedly masked and communicated for each secure dot product. We observe that the sparsity distribution exhibits strong spatial locality, meaning that dot products involving nonzero outputs or inputs in the same row or column often use the same data. Building on this spatial locality, we propose the \emph{Sparse Output Matrix Multiplication} (SOMM) protocol for the preceding linear layer and the \emph{Sparse Input Matrix Multiplication} (SIMM) protocol for the subsequent linear layer to minimize communication and computation. 

Since the sparsity distribution output by the predictor has been shuffled, we also perform an oblivious shuffle on the input matrix before executing SOMM protocol. This aligns the input matrix permutation with the shuffled sparsity distribution, allowing secure and correct indexing of its rows, columns, or elements. Notably, since the model weight matrices are typically fixed, we can perform a one-time, offline pre-shuffle on them, which does not add any latency during online inference. After executing SIMM protocol, we perform an unshuffle on the output matrix to restore it to its original order, ensuring correct computation for the next layer. Without loss of generality, in the following discussion, we assume that the sparsity distribution and all matrices are in the same permutation.

\subsection{Sparse output matrix multiplication protocol for preceding linear layer}
\label{subsec:SOMM}

\begin{figure}[htbp]
    \centering
    \includegraphics[width=0.45\textwidth]{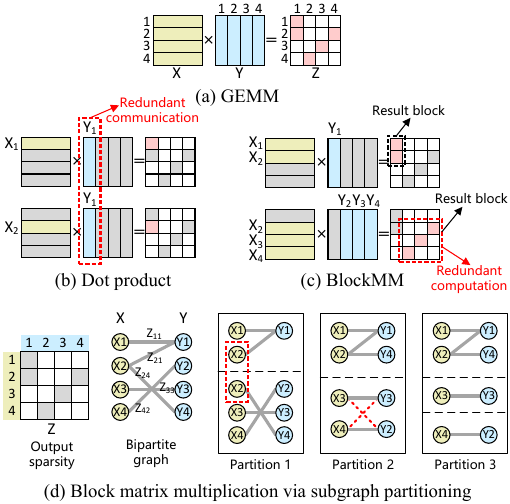}
    \caption{Sparse output matrix multiplication.}
    \label{fig:SOMM}
    \vspace{-0.5cm}
\end{figure}

For the preceding linear layer matrix multiplication $\mathbf{X} \mathbf{Y} = \mathbf{Z}$, given the sparsity distribution of $\mathbf{Z}$,
we can pre-identify which output elements need to be computed (shown in pink in Figure~\ref{fig:SOMM}(a)), and perform dot products only for relevant rows and columns (shown in Figure~\ref{fig:SOMM}(b)).
In MPC, the dot product is computed using a Beaver triple $(\shareB{a}, \shareB{b}, \shareB{c})$. To calculate $\share{\bZ_{i,j}}$ , the MPC servers first locally compute $\share{\bE_i} = \share{\bX_i} - \shareB{a}$ and $\share{\bF_j} = \share{\bY_j} - \shareB{b}$, then exchange these masked values with other servers, and finally derive $\share{\bZ_{i,j}}$ using Equation~\ref{eq:multiplication}. 
Due to security requirements, if $\share{\bX_i}$ or $\share{\bY_j}$ is involved in multiple separate dot products, each instance must use a fresh Beaver triple~\cite{beaver1992efficient}. As a result, when the same data participates in multiple separate dot products, it leads to repeated masking and communication. For instance, $\bY_1$ in Figure~\ref{fig:SOMM}(b) is involved in two dot products, requiring masking and communication twice.

Based on the spatial locality of the sparsity distribution, we group nonzero ouputs computed from the same row or column into a single result block. This allows us to perform block matrix multiplications, where each row or column is masked and communicated only once in MPC. As shown in Figure~\ref{fig:SOMM}(c), by grouping $\bZ_{1,1}$ and $\bZ_{2,1}$, the MPC servers perform matrix multiplicaiton $\Pi_\textsf{MatMul}(\share{\bX_1, \bX_2}, \share{\bY_1})$ instead of dot products $\Pi_\textsf{DP}(\share{\bX_1}, \share{\bY_1})$ and $\Pi_\textsf{DP}(\share{\bX_2}, \share{\bY_1})$, where $\share{\bY_1}$ only needs to be masked and communicated once. However, naive grouping may introduce redundant computations, as these result blocks might include zero outputs, leading to unnecessary operations, as illustrated in the lower part of Figure~\ref{fig:SOMM}(c).

To determine the grouping with optimal communication and computation, we model it as a subgraph partition problem in a bipartite graph.
As shown in Figure~\ref{fig:SOMM}(d), the matrix multiplication is represented as a bipartite graph, where rows of $\mathbf{X}$ and columns of $\mathbf{Y}$ are nodes, and the nonzero elements of the output $\bZ$ are edges. The number of nodes represents the communication cost (the number of rows and columns need to be masked and communited), while the product of nodes reflects the computation cost (the number of dot products). Grouping nonzero elements of $\bZ$ into result blocks corresponds to partitioning the bipartite graph into subgraphs. To achieve optimal partitioning, two requirements must be met:

\begin{itemize}[leftmargin=1em]
    \item Minimize communication cost: Each connected component should be fully contained within a single subgraph. If a node appears in multiple subgraphs, the corresponding row or column will be involved in multiple individual matrix multiplications, causing redundant communication. For example, in Figure~\ref{fig:SOMM}(d), adjusting edge $\bZ_{2,4}$ and its nodes into partition 2 prevents redundant communication of node $\bX_2$ from partition 1.
    \item Optimize computation: Each subgraph should contain only one connected component. If multiple components exist in a subgraph, redundant computations are introduced. These redundancies can be avoided without increasing communication costs, as shown by partitioning the two connected components in partition 2 into partition 3 in Figure~\ref{fig:SOMM}(d).
\end{itemize}

Therefore, finding the optimal partition is equivalent to identifying all connected components in the bipartite graph, which can be quickly accomplished using depth-first search (DFS)~\cite{tarjan1972depth}. This only needs to know the positions of nonzero outputs, and can be performed locally on each MPC server, without communication.

\begin{algorithm}[h]
    \caption{SOMM Protocol $\Pi_{\mathsf{SOMM}}$}
    \label{protocol:SOMM}
    \begin{algorithmic}[1]
        \Input
        $P_i$ holds the share $\share{\mathbf{X}}_i$, the share $\shareB{Y}_i$ and the sparsity distribution $\mathcal{S}$ of $\mathbf{X}\mathbf{Y}$, $i\in \{1,2\}$.
        \Output
        $P_i$ gets the share $\share{\mathbf{Z}}_i$, such that $\mathbf{Z}=\mathbf{X}\mathbf{Y}$.
        \State $P_i$ locally computes $\{G_t\}_{t=1}^L \gets \mathsf{DFS}(\mathcal{S})$.
        \For{$t=1$ to $L$ parallel}
            \State $P_i$ locally transforms the graph $G_t$ into row indices set $I_x$ and column indices set $I_y$.
            \State $P_i$ locally extracts $\shareB{X}_i^t \gets \mathsf{IndexRows}(\shareB{X}_i, I_x)$.
            \State $P_i$ locally extracts $\shareB{Y}_i^t \gets \mathsf{IndexCols}(\shareB{Y}_i, I_y)$.
            \State $P_i$ invokes $\shareB{R}_i^t \gets \Pi_\mathsf{MatMul}(\shareB{X}_i^t, \shareB{Y}_i^t)$.
        \EndFor
        \State $P_i$ locally computes $\shareB{Z}_i \gets \mathsf{Merge}(\{\shareB{R}_i^t\}_{t=1}^{L})$.
        \State \Return $\share{\mathbf{Z}}_i$
    \end{algorithmic}
\end{algorithm}

Based on the above analysis, we propose the SOMM protocol for the preceding linear layer. The detailed steps are provided in Protocol~\ref{protocol:SOMM}. First, each party locally performs a depth-first search on the sparsity distribution to identify all connected components (line 1). Then, these connected components are transformed into corresponding index sets (line 3), which are used to extract the respective submatrices (line 4 and 5). Subsequently, a secure matrix multiplication protocol is invoked on these submatrices individually to obtain the result block (line 6). Finally, all the result blocks are merged to form the final output (line 8). The security of this protocol is proven in Appendix~\ref{sec:security_analysis}. The proof of Theorem~\ref{theorem:SOMM} can be found in Appendix~\ref{sec:appendix_somm}.

\begin{theorem}
    \label{theorem:SOMM}
    Given $\shareB{X}$, $\shareB{Y}$, and the sparsity distribution $\mathcal{S}$ of $\mathbf{XY}$, Protocol~\ref{protocol:SOMM} achieves the minimal communication cost for computing $\shareB{XY}$. Furthermore, under this minimal communication cost, the protocol also ensures the minimal computation cost.
\end{theorem}

\subsection{Sparse input matrix multiplication protocol for subsequent linear layer}
\label{subsec:SIMM}

\begin{figure}[htbp]
    \centering
    \includegraphics[width=0.45\textwidth]{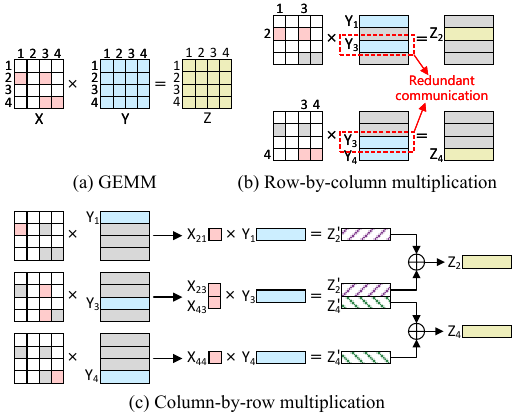}
    \caption{Sparse input matrix multiplication.}
    \label{fig:SIMM}
\end{figure}

For the subsequent linear layer matrix multiplication $\mathbf{X} \mathbf{Y} = \mathbf{Z}$, given the sparsity distribution of $\mathbf{X}$, this operation can be seen as a generalized sparse matrix-matrix multiplication (SpGEMM). Efficient implementations already exist in plaintext computation, such as the PyTorch sparse tensor library~\cite{pytorch_sparse_mm}. A typical SpGEMM approach follows a row-by-column multiplication: each nonzero element in $\mathbf{X}$ is multiplied with the corresponding elements in the respective column of $\mathbf{Y}$. Specifically, for a nonzero $\mathbf{X}_{i,j}$, it suffices to multiply it with the $j$-th row of $\mathbf{Y}$, as shown in Figure~\ref{fig:SIMM}(b).
However, in MPC, when multiple nonzero elements exist in the same column of $\shareB{X}$, the corresponding rows in $\shareB{Y}$ must participate in multiple separate computations, which leads to redundant communications.
For example, in Figure~\ref{fig:SIMM}(b), the third column of $\bX$ contains two nonzero elements, $\bX_{2,3}$ and $\bX_{4,3}$. As a result, $\bY_3$ is involved in two separate multiplications, requiring $\share{\bY_3}$ to be masked and communicated twice in MPC.

To avoid this redundant communication, we shift from row-by-column multiplication to column-by-row multiplication. Nonzero elements in the same column of $\mathbf{X}$ are aggregated into a sub-vector, which is then multiplied with the corresponding row of $\mathbf{Y}$, as shown in Figure~\ref{fig:SIMM}(c). This ensures that each row of $\shareB{Y}$ participates in the multiplication only once in MPC, requiring masking and communication only once as well.
After performing the multiplication for each sub-vector, the results belonging to the same row of the output $\mathbf{Z}$ are summed and merged to obtain the final results (e.g., shaded vectors in Figure~\ref{fig:SIMM}(c) add up to $\mathbf{Z}_2$ and $\mathbf{Z}_4$).

\begin{algorithm}[h]
    \caption{SIMM Protocol $\Pi_{\mathsf{SIMM}}$}
    \label{protocol:SIMM}
    \begin{algorithmic}[1]
        \Input
        $P_i$ holds the share $\shareB{X}_i$, the share $\shareB{Y}_i$, the sparsity distribution $\mathcal{S}$ of $\mathbf{X}$, $i\in \{1,2\}$.
        \Output
        $P_i$ gets the share $\share{\mathbf{Z}}_i$, such that $\mathbf{Z}=\mathbf{X}\mathbf{Y}$.
        \State $I \gets \varnothing$, $P_i$ finds all nonzero columns $\{j\}$ in $\mathcal{S}$.
        \For{each nonzero column $j$}
            \State $P_i$ finds all nonzero row indices set $R_j$.
            \State $I \gets I \cup \{(j, R_j)\}$
        \EndFor
        \For{$(j, R_j)$ in $I$ parallel}
            \State $P_i$ locally extracts $\shareB{x}'_i \gets \mathsf{IndexCol}(\shareB{X}_i, j)$.
            \State $P_i$ locally extracts $\shareB{x}''_i \gets \mathsf{IndexRows}(\shareB{x}'_i, R_{j})$.
            \State $P_i$ locally extracts $\shareB{y}_i \gets \mathsf{IndextRow}(\shareB{Y}_i, j)$.
            \State $P_i$ invokes $\shareB{R}_i^t \gets \Pi_\mathsf{MatMul}(\shareB{x}''_i, \shareB{y}_i)$.
        \EndFor
        \State $P_i$ locally computes $\shareB{Z}_i \gets \mathsf{Merge\&Sum}(\{\shareB{R}_i^t\}_{t=1}^{L})$.
        \State \Return $\share{\mathbf{Z}}_i$
    \end{algorithmic}
\end{algorithm}

Based on the above analysis, we propose the SIMM protocol for the subsequent linear layer. The detailed steps are provided in Protocol~\ref{protocol:SIMM}. First, each party finds all nonzero columns in $\mathcal{S}$ and records the nonzero row indices within each column (line 1-5). For each nonzero column of $\shareB{X}$, each party extracts its nonzero elements to form a sub-vector(line 7 and 8) and use the column index to extract the corresponding row of $\shareB{Y}$ (line 9). A secure matrix multiplication protocol is then invoked to compute the result block for the sub-vector and the row (line 10). Finally, all result blocks corresponding to the same row are summed, and these rows are combined to form the final output matrix (line 12). The security of this protocol is proven in Appendix~\ref{sec:security_analysis}. The proof of Theorem~\ref{theorem:SIMM} can be found in Appendix~\ref{sec:appendix_simm}.

\begin{theorem}
    \label{theorem:SIMM}
    Given $\shareB{X}$, $\shareB{Y}$, and the sparsity distribution $\mathcal{S}$ of $\mathbf{X}$, Protocol~\ref{protocol:SIMM} achieves the minimal communication cost for computing $\shareB{XY}$. Furthermore, under this minimal communication cost, the protocol also ensures the minimal computation cost.
\end{theorem}

%% file: chapters/6_kv_cache.tex
\section{KV Cache Manager}

During inference, LLMs take the user's prompt as input and generate text token by token. For each token generation, the MHA requires accessing the keys(K) and values(V) of all previous tokens to compute attention scores. Existing LLM systems employ a KV cache to store these data for efficient reuse.

However, when some attention heads are predicted to be not activated, the Sparse QKV layer skips the computation of their keys and values, leaving these entries absent from the KV cache. This disrupts the temporal (token-wise) and spatial (attention head-wise) continuity of the KV cache, leading to a reduced hit rate. 
If these missing key-value pairs are ignored in subsequent computations, model accuracy may degrade. To mitigate this, we need to refill in the missing cache entries upon detection. However, plaintext cache refill strategies~\cite{li2024survey} are designed to minimize computation cost, and directly applying them to priavete inference can introduce substantial communication overhead.

To address these challenges, we propose two complementary strategies: merging cache miss requests to reduce redundant computation and communication, and prefetching KV values for not-activated attention heads to improve cache continuity.

\begin{figure}[htbp]
    \centering
    \includegraphics[width=0.48\textwidth]{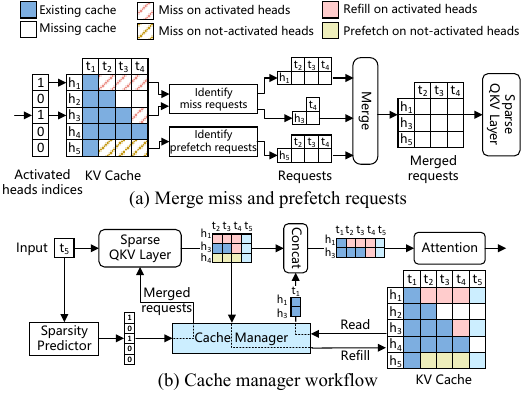}
    \caption{KV cache manager.}
    \label{fig:kv_cache}
    \vspace{-0.5cm}
\end{figure}

\subsection{Merging cache miss requests}
A straightforward approach to handle cache misses is to respond to each miss request immediately by refilling the cache entry. This requires the MPC servers to collaboratively execute matrix multiplications in the Sparse QKV layer. However, when multiple requests target the same attention head, separatly processing each request results in redundant communication, as the QKV weights must be repeatedly communicated and masked, causing significant latency.

To mitigate this issue, we propose merging cache miss requests before performing secure matrix multiplications. As illustrated in Figure~\ref{fig:kv_cache}(a), when generating a token, attention heads $h_1$ and $h_3$ are predicted to be activated. The cache manager identifies the cache misses for these heads (shaded in pink) and merges the requests into a single batch. The Sparse QKV layer then processes the merged requests together, ensuring that each token and attention head participates in the computation only once, reducing redundant communication.

\subsection{Prefetching key-value for not-activated heads}

While merging addresses immediate cache misses, prefetching proactively improves the KV cache continuity for future tokens. Specifically, when a cache miss occurs for an activated attention head, the cache manager triggers a prefetching mechanism to compute and store the KV values for some not-activated attention heads (shaded in yellow in Figure~\ref{fig:kv_cache}(a)). By leveraging shared communication for the same tokens, prefetching reduces future cache misses and associated latency.

The key challenge is determining which not-activated attention heads are worth prefetching. We propose a cost-benefit-based head selection strategy, prefetching only when \textit{the future communication savings from reduced cache misses} outweigh \textit{the immediate communication cost incurred by performing the prefetch}. Let $x$ denote the token vector size and $w$ the size of the attention head weight matrix. For a prefetched not-activated head with $L_2$ cache misses (while there are $L_1$ misses for activated heads), the additional communication cost is $2(w + x \cdot \max(0, L_2 - L_1))$, while the saved cost is $2L_2x$. Prefetching is executed only if $L_2 > w/x + \max(0, L_2 - L_1)$, ensuring cost-effectiveness. For instance, in OPT-6.7B, at least 128 cache misses are required for a not-activated head to be selected for prefetching.

Finally, because refilling and prefetching requests often share tokens, they are merged into a single batch (Figure~\ref{fig:kv_cache}(a)). Moreover, as these requests use the same activated attention heads as the current token's inference, they are further combined and processed in the Sparse QKV Layer (Figure~\ref{fig:kv_cache}(b)). Within this layer, keys and values are computed using \(\Pi_\textsf{shuffle}\) and \(\Pi_\textsf{SOMM}\), so they are already shuffled in the KV cache. Consequently, neither refilling nor prefetching reveals any information: the server cannot tell which attention heads are missing or have been prefetched.

%% file: chapters/8_evaluation.tex
\begin{figure*}[h]
    \centering
    \includegraphics[width=0.95\textwidth]{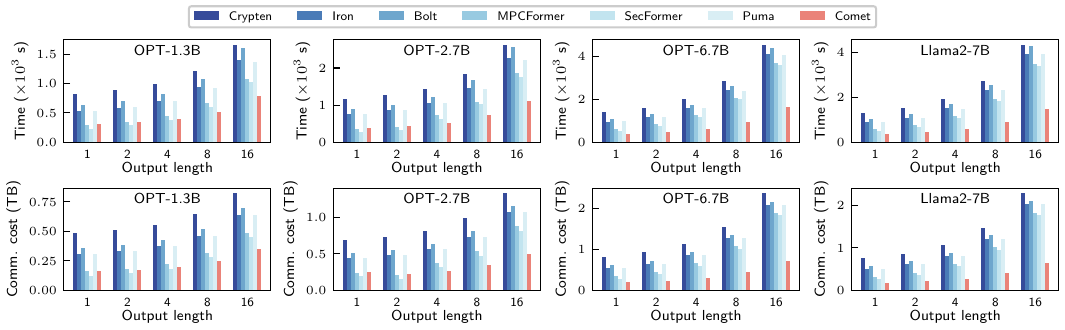}
    \caption{Overall performance of {\projectname}.}
    \label{fig:overall_performance}
     \vspace{-0.3cm}
\end{figure*}

\section{Evaluation}
\label{sec:evalution}
\subsection{Methodology}

\mypara{Hardware platform.} We evaluate {\projectname} on three cloud servers, with two servers for MPC computation node and one for user node.
Each server is equipped with an Intel 8458P@2.7GHz CPU, an NVIDIA A100 GPU, and 256GB of memory. The bandwidth between the servers is 5Gbps.

\mypara{Software platform.} {\projectname} is implemented based on Crypten~\cite{knott2021crypten}, a widely-used MPC framework. By leveraging MPC primitives provided by Crypten, we implement the private inference protocol in {\projectname}, including oblivious shuffle, SOMM and SIMM. The activation sparsity predictor and cache manager are designed as independent modules, without modification to the model architecture. 

\mypara{Models}
We use two popular LLMs, namely OPT~\cite{zhang2022opt} with parameters from 1.3B to 6.7B and LLaMA2(ReLU)-7B~\cite{song2024prosparse}. 
The model pre-trained weights are sourced from the Transformer library HuggingFace~\cite{wolf2019huggingface}, which provides a variety of pre-trained models.

\mypara{Workloads.} 
The workloads for experiments are derived
from Alpaca~\cite{taori2023stanford} datasets, which consists of input and output texts typical of real LLM services. To test the accuracy, we used eight popular LLM benchmarks, covering four task types. (1) Code Generation: HumanEval (0-shot)~\cite{chen2021evaluating} and MBPP (3-shot)~\cite{austin2021program}. (2) Commonsense Reasoning: PIQA (0-shot)~\cite{bisk2020piqa} and COPA (0-shot)~\cite{roemmele2011choice}. (3) Reading Comprehension: BoolQ (0-shot)~\cite{clark2019boolq} and LAMBADA (0-shot)~\cite{paperno2016lambada}.  (4) Language Understanding: MMLU (5-shot)~\cite{hendrycks2020measuring} and GLUE~\cite{wang2018glue}.

\mypara{Baselines.} 
We select five mainstream Transformer private inference systems as baselines for comparison with {\projectname}. Two systems, Iron~\cite{hao2022iron} and Bolt~\cite{pang2023bolt}, use hybrid protocols with homomorphic encryption (HE) for linear layers and MPC for non-linear layers. Since HE is slower than MPC, we ensure a fair comparison by using their non-linear layer implementations, while keeping the linear layer implementation consistent with {\projectname}. The other three baselines, MPCFormer~\cite{li2022mpcformer}, SecFormer~\cite{luo2024secformer}, and Puma~\cite{dong2023puma}, rely solely on MPC. We provide detailed descriptions of the baselines in Appendix~\ref{sec:appendix_baselinse}.

\mypara{Metrics.} Our metrics include end-to-end runtime and communication cost. The end-to-end runtime measures the total time  reflecting the latency of private inference. We quantify the communication cost as the total amount of data exchanged by the MPC servers during inference, mitigating the impact of network fluctuation.

\subsection{End-to-End Performance}

We evaluate the end-to-end performance of the {\projectname} for generating text of different lengths (1, 2, 4, 8, and 16), using an input length of 512 tokens.
The results are presented in Figure~\ref{fig:overall_performance}. 
Compared to the baselines, {\projectname} achieve faster inference speeds ($1.87\times$ to $2.63\times$) and lower communication overhead ($1.94 \times$ to $2.64\times$).

{\projectname} is effective for models of varying sizes, exhibiting more pronounced speedups on larger models. For the 1.3B model (OPT-1.3B), {\projectname} achieves an average speedup of $1.71\times$, while reducing communication cost by $1.61\times$. When the model size increases to 7B (Llama2-7B), the average speedup of {\projectname} further increases to $2.58\times$, with a reduction in communication cost of $2.49\times$. This is because the larger models typically present higher activation sparsity, which results in more redundant computations and communication, thereby offering greater potential for acceleration. For instance, the average sparsity of the MHA and FFN in OPT-1.3B is $34\%$ and $61\%$, respectively, while Llama2-7B reaches $49\%$ and $85\%$.

As the output length increases, the speedup improves due to the two-stage inference process: Prefill and Decode. For instance, with the OPT-6.7B model, when generating the first token (Prefill), {\projectname} achieves a speedup of $2.09\times$ and reduces communication cost by $2.14\times$. However, when generating the sixteenth token (Decode), the total speedup increases to $2.57\times$ and the total communication cost is reduced by $2.73\times$. 
This is because, during the Decode stage, only one token is input at a time, leading to high sparsity. As the sequence length increases, this stage increasingly dominates the inference latency, leading to greater speedup and reduced communication costs.

We also evaluated the end-to-end performance of {\projectname} under different bandwidths, from 100Mbps to 5Gbps, as shown in Table~\ref{tab:bandwidth_comparsion}. We only selected two faster methods (MPCFormer and Puma) on the two largest models (OPT-6.7B and Llama2-7B) for comparison. Under lower bandwidth (100Mbps, WAN), {\projectname} achieved a $3.25\times$ average speedup, and under higher bandwidth (5Gbps, LAN), it achieved a $2.05\times$ average speedup. Overall, {\projectname} accelerates private inference across all bandwidth conditions by directly reducing communication costs.

\begin{table}[]
    \scriptsize
    \caption{End-to-end performance of {\projectname} for generating one token with input length 512 under different bandwidths.}
    \centering
    \begin{tabular}{ccccccc}
         \toprule
          \multirow{2}{*}{\textbf{Model}}
          & \multirow{2}{*}{\textbf{Method}}
          & \multicolumn{4}{c}{\textbf{Time} (min)}
          \\ \cmidrule{3-6}
          & & 100Mbps & 500Mbps & 1Gbps & 5Gbps\\
          \midrule
          \multirow{3}{*}{\rotatebox{0}{OPT-6.7B}}
               & MPCFormer & 530.2 & 107.6 & 48.9 & 10.6\\
               & Puma & 752.1 & 157.9 & 76.4 & 16.1 \\
               & \projectname & \textbf{201.8} & \textbf{58.3} & \textbf{27.7} & \textbf{7.7}\\
          \midrule
          \multirow{3}{*}{\rotatebox{0}{Llama2-7B}} 
               & MPCFormer & 575.5 & 98.4 & 44.0 & 9.5\\
               & Puma & 722.7 & 154.5 & 70.4 & 14.9\\
               & \projectname & \textbf{194.8} & \textbf{54.1} & \textbf{25.8} & \textbf{7.0}\\
         \bottomrule
    \end{tabular}
    \label{tab:bandwidth_comparsion}
\end{table}

\subsection{Performance Breakdown}

To analyze the acceleration effect of {\projectname} on linear and non-linear layers, we conduct a detailed breakdown. The experiments use Puma as the baseline due to its superior performance.

\begin{table*}[ht]
    \centering
    \scriptsize
    \caption{The speedup of {\projectname} for different layers. The input length is 512 and output length is 16.}
    \begin{tabular}{p{0.1em}c|P{0.8cm}P{0.8cm}P{0.8cm}P{0.95cm}|P{0.8cm}P{1cm}P{0.8cm}P{1cm}P{0.8cm}P{1cm}P{0.8cm}P{1cm}}
         \toprule
              \multirow{2}{*}{}  
              &\multirow{3}{*}{\textbf{Layer}}
              & \multicolumn{4}{c|}{\textbf{PUMA}}
              & \multicolumn{8}{c}{\textbf{\projectname}}
              \\
              & & Comp. & Comm. & Total & Comm. & \multicolumn{2}{c}{Comp.} & \multicolumn{2}{c}{Comm.} & \multicolumn{2}{c}{Total} & \multicolumn{2}{c}{Comm.}
              \\
              & & Time(s) & Time(s) & Time(s) & Cost(GB) & Time(s) & Speedup & Time(s) & Speedup & Time(s) & Speedup & Cost(GB) & Reduce
              \\ \hline
              \multirow{9}{*}{\rotatebox{90}{OPT-6.7B}}
              & QKV & 78.5 & 618 & 697 & 387 & 53.0 & $1.48\times$ & 309 & $2.0\times$ & 362 & $1.92\times$ & 191 & $2.02\times$ 
              \\
              & MatMul & 10.3 & 60.2 & 70.5 & 37.6 & 7.3 & $1.41\times$ & 29.1 & $2.07\times$ & 36.4 & $1.94\times$ & 17.5 & $2.15\times$ 
              \\
              & Softmax & 221 & 524 & 745 & 330 & 185 & $1.19\times$ & 230 & $2.27\times$ & 416 & $1.79\times$ & 141 & $2.34\times$ 
              \\
              & Output & 25.7 & 207 & 232 & 129 & 9.8 & $2.63\times$ & 99.6 & $2.08\times$ & 109 & $2.13\times$ & 59.1 & $2.18\times$ 
              \\
              & FC1 & 100 & 849 & 949 & 513 & 30.6 & $3.28\times$ & 85.0 & $10.0\times$ & 115 & $8.22\times$ & 53.2 & $9.65\times$ 
              \\
              & ReLU & 13.4 & 189 & 203 & 119 & 10.3 & $1.3\times$ & 16.1 & $11.8\times$ & 26.4 & $7.7\times$ & 10.1 & $11.9\times$ 
              \\
              & FC2 & 100 & 816 & 916 & 516 & 13.6 & $7.37\times$ & 83.5 & $9.78\times$ & 97.1 & $9.44\times$ & 52.6 & $9.81\times$ 
              \\
              & Others & 94.4 & 171 & 265 & 106 & 95.7 & $0.99\times$ & 169 & $1.01\times$ & 265 & $1.0\times$ & 106 & $1.0\times$ 
              \\
              & Predictor & 0 & 0 & 0 & 0 & 28.7 & 0 & 168 & 0 & 196 & 0 & 104 & 0 
              \\
              & \textbf{Total} & 644 & 3437 & 4082 & 2140 & \textbf{434} & \textbf{1.48}$\times$ & \textbf{1190} & \textbf{2.89}$\times$ & \textbf{1625} & \textbf{2.51}$\times$ & \textbf{742} & \textbf{2.88}$\times$ 
              \\ \hline
              \multirow{10}{*}{\rotatebox{90}{Llama2-7B}} 
              & QKV & 75.9 & 618 & 694 & 387 & 50.9 & $1.49\times$ & 293 & $2.11\times$ & 343 & $2.02\times$ & 180 & $2.14\times$ 
              \\
              & MatMul & 10.3 & 60.1 & 70.4 & 37.6 & 7.6 & $1.35\times$ & 26.8 & $2.24\times$ & 34.4 & $2.04\times$ & 16.5 & $2.28\times$ 
              \\
              & Softmax & 224 & 529 & 753 & 330 & 175 & $1.28\times$ & 211 & $2.5\times$ & 387 & $1.95\times$ & 133 & $2.48\times$ 
              \\
              & Output & 24.7 & 205 & 230 & 129 & 8.9 & $2.78\times$ & 92.5 & $2.23\times$ & 101 & $2.27\times$ & 55.7 & $2.32\times$ 
              \\
              & Gate & 66.4 & 548 & 614 & 345 & 21.6 & $3.07\times$ & 57.7 & $9.51\times$ & 79.3 & $7.75\times$ & 36.1 & $9.57\times$ 
              \\
              & Up & 66.3 & 557 & 623 & 345 & 19.9 & $3.33\times$ & 57.2 & $9.74\times$ & 77.1 & $8.09\times$ & 36.1 & $9.57\times$ 
              \\
              & ReLU & 13.4 & 131 & 145 & 80.2 & 11.6 & $1.15\times$ & 12.4 & $10.6\times$ & 24.1 & $6.03\times$ & 7.9 & $10.2\times$ 
              \\
              & Down & 64.5 & 547 & 611 & 346 & 8.9 & $7.21\times$ & 56.2 & $9.74\times$ & 65.1 & $9.39\times$ & 35.4 & $9.79\times$ 
              \\
              & Others & 87.4 & 113 & 200 & 70.3 & 85.3 & $1.02\times$ & 112 & $1.0\times$ & 198 & $1.01\times$ & 70.3 & $1.0\times$ 
              \\
              & Predictor & 0 & 0 & 0 & 0 & 22.7 & 0 & 130 & 0 & 153 & 0 & 81.4 & 0 
              \\
              & \textbf{Total} & 633 & 3311 & 3945 & 2072 & \textbf{413} & \textbf{1.53}$\times$ & \textbf{1051} & \textbf{3.15}$\times$ & \textbf{1464} & \textbf{2.69}$\times$ & \textbf{659} & \textbf{3.14}$\times$ 
              \\
         \bottomrule
    \end{tabular}
    \label{tab:performance_breakdown}
\end{table*}

\begin{figure}[h]
    \centering
    \includegraphics[width=0.42\textwidth]{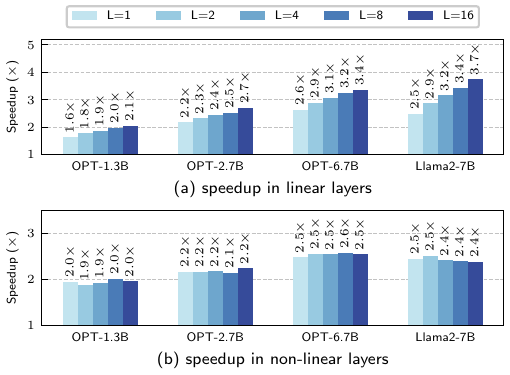}
    \caption{Speedup breakdown for linear and non-linear layers across different output lengths (denoted as $L$).}
    \label{fig:breakdown_linear_non-linear}
\end{figure}

As shown in Figure~\ref{fig:breakdown_linear_non-linear}, {\projectname} accelerates both linear and non-linear layers, with speedups of $1.6\times$ to $3.7\times$ for linear layers and $1.9\times$ to $2.6\times$ for non-linear layers. Linear layer speedup increases with sequence length due to growing sparsity, while non-linear layers are less affected. In linear layers, sparsity is low when generating the first token, leading to modest acceleration, but increases for subsequent tokens, boosting acceleration further. For non-linear layers, the time for each token decreases significantly after the first, so the acceleration mainly depends on the first token. For example, in OPT-6.7B, the linear layer's communication overhead rises from $24\%$ for the first token to $75\%$ by the sixteenth.

We evaluate the performance for each model layer to gain a deeper understanding. Table~\ref{tab:performance_breakdown} presents the results for the two largest models, OPT-6.7B and Llama2-7B. All layers are accelerated,
and the performance improvement in the layers within the FFN (FC1, ReLU, FC2) is more significant than that in the layers within the MHA (QKV, MatMul, Softmax, Output). 
Specifically the MHA and FFN achieve an average speedup of $1.94\times$ and $8.08\times$, and a communication reduction of $2.23\times$ and $10.01\times$, respectively.
This is due to the higher sparsity rate of the FFN (approximately 50\%) compared to that of the MHA (about 90\%).

\subsection{Accuracy}
\label{subsec:accuracy}
We evaluated the accuracy of {\projectname} on different datasets using Llama2-7B. The sparsity predictor was trained on the WikiText2~\cite{merity2016pointer} and StarCoder~\cite{li2023starcoder} datasets, which are different from the inference datasets. This ensures that the accuracy is not influenced by prior exposure to the inference data. As shown in Table~\ref{tab:generality_tasks}, {\projectname} maintains accuracy comparable to plaintext inference, with a average accuracy loss of 1.5\%. The predictor achieves an average recall of 93\%.

\begin{table}[H]
    \scriptsize
    \centering
    \caption{Accuracy of {\projectname} under different tasks.}
    \begin{tabular}{ccc|ccc}
        \toprule \textbf{Dataset} & \textbf{Plaintext} & \textbf{{\projectname}} & \textbf{Dataset} & \textbf{Plaintext} & \textbf{{\projectname}}
        \\
        \midrule
        HumanEval & 18.8 & 18.2 & MBPP & 22.4 & 91.2
        \\
        PIQA & 78.4 & 76.8 & COPA & 81.3 & 78.5
        \\
        BoolQ & 62.5 & 60.4 & Lambada & 66.2 & 63.4
        \\
        MMLU & 44.5 & 42.8 & GLUE & 84.7 & 83.5
        \\
        \bottomrule
    \end{tabular}
    \label{tab:generality_tasks}
\end{table}

\begin{figure}[h]
    \centering
    \includegraphics[width=0.38\textwidth]{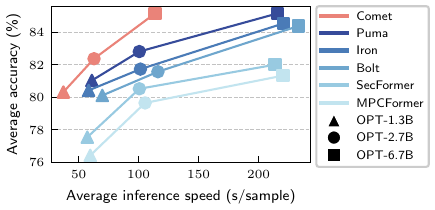}
    \caption{Accuracy and inference speed of various methods on the GLUE benchmark.}
    \label{fig:accuracy}
\end{figure}

{\projectname} achieves a better trade-off between accuracy and inference speed compared to other methods. As shown in Figure~\ref{fig:accuracy}, for larger models (OPT-6.7B), {\projectname} achieves approximately $2.1\times$ speedup over Puma while maintaining the same accuracy (85.17\%). For smaller models (OPT-1.3B), although accuracy drops slightly (about 0.7\% lower than Puma and Iron), {\projectname} still delivers a notable speedup ($1.6\times$ faster than Puma). As model size increases, {\projectname} further accelerates inference while preserving accuracy.

\subsection{Ablation Study}

\begin{figure}[h]
    \centering
    \includegraphics[width=0.42\textwidth]{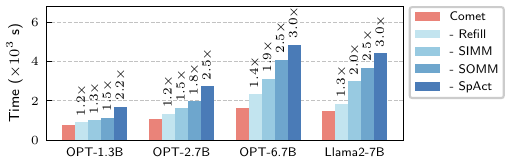}
    \caption{Ablation study.}
    \label{fig:ablation}
    \vspace{-0.3cm}
\end{figure}

To demonstrate the effectiveness of each design within the {\projectname}, we conduct an ablation study. Figure~\ref{fig:ablation} illustrates the inference time as each component is progressively removed. Initially, removing the cache refilling strategy results in a $1.2\times$ to $1.4\times$ increase in inference time. Further removal of SIMM leads to a rise in inference latency by $1.3\times$ to $2.0\times$, and the subsequent removal of SOMM increases inference time by $1.5\times$ to $2.5\times$. 
Finally, removing the predictor (denoted as "-SpAct") caused the system to revert to a standard method without any optimizations. 
Overall, all the designs have proven to be effective.

\subsection{Component-wise Analysis}

To provide a deeper understanding of the contributions of individual components in {\projectname}, we perform a detailed component-wise analysis. Each subsection focuses on a specific design element, evaluating its unique characteristics and effectiveness in addressing different aspects of private inference.

\mypara{(1) Activation sparisity predictor.} 

\begin{figure}[h]
    \centering
    \includegraphics{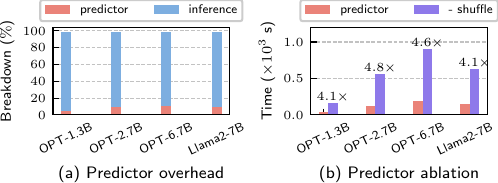}
    \caption[width=0.4\textwidth]{Predictor overhead.}
    \label{fig:ablation_overhead_of_predictor}
\end{figure}

\vspace{2pt}\emph{Prediction overhead.} 
As shown in Figure~\ref{fig:ablation_overhead_of_predictor}(a), for models of different sizes, the average overhead of the predictor does not exceed 15\% of the entire end-to-end inference time. This low prediction overhead is due to our oblivious shuffle design. As shown in Figure~\ref{fig:ablation_overhead_of_predictor}(b), the oblivious shuffle design brings a $4.1\times$ to $4.8\times$ performance improvement. 

\begin{figure}[h]
    \centering
    \includegraphics[width=0.4\textwidth]{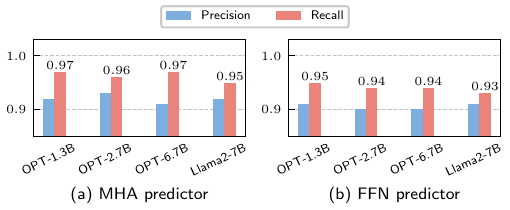}
    \caption{Precision and recall of predictor.}
    \label{fig:ablation_precision_recall_of_predictor}
    \vspace{-0.3cm}
\end{figure}

\vspace{2pt}\emph{Predictor accuracy.} 
We evaluate the precision and recall of the predictor on models of varying sizes. A lower precision indicates that more not activated neuron are incorrectly predicted as activated, introducing additional avoidable computations and communication. Conversely, a lower recall suggests that more activated neurons are missed, thereby impacting the model's accuracy. As illustrated in Figure~\ref{fig:ablation_precision_recall_of_predictor}, the predictor achieved an average precision of up to $90\%$ and a recall of $95\%$. Although the precision is slightly lower then the recall, it only result in performance loss.

\vspace{2pt}\emph{Predictor threshold.}
We analyze how different predictor thresholds affect {\projectname}'s accuracy and inference speed using the Llama2-7B model. During pretraining, the predictor threshold was set to 0. We tested various thresholds on the GLUE benchmark. As shown in Table~\ref{tab:threshold_performance}, keeping the same threshold as in pretraining yields the highest accuracy by preserving most neuron computations (89.1\% sparsity). Increasing the threshold filters out more neurons, leading to a faster inference speed but a sharp drop in accuracy. In most tasks, maintaining the pretraining threshold provides a good balance between accuracy and efficiency.

\begin{table}[h]
    \centering
    \caption{Impact of predictor thresholds on {\projectname}'s accuracy and speed.}
    \label{tab:threshold_performance}
    \begin{tabular}{cccccc}
        \toprule
        \textbf{Threshold}  & 0  & 0.1  & 0.3  & 0.5  & 0.7  \\
        \midrule
        \textbf{Sparsity} (\%)  & 89.1 & 92.3 & 94.8 & 96.7 & {99.1} \\
        \textbf{Accuracy} (\%) & {86.4} & 81.5 & 64.3 & 53.1 & 49.3 \\
        \textbf{Speedup}        & 2.7$\times$ & 2.9$\times$ & 3.4$\times$ & 4.4$\times$ & {5.7$\times$} \\
        \bottomrule
    \end{tabular}
\end{table}

\begin{figure}[h]
    \centering
    \includegraphics{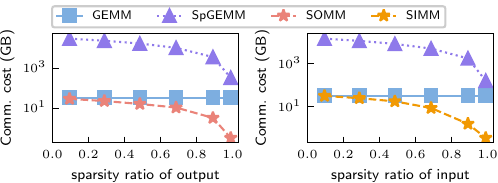}
    \caption{Communication cost comparison of matrix multiplication methods at different sparsity levels.}
    \label{fig:ablation_spmm}
\end{figure}

\mypara{(2) Sparse matrix multiplication.}
We compare the performance of SOMM, SIMM, GEMM, and SpGEMM under different sparsity levels, focusing on communication costs since computation time is only about $15\%$ of the total. The input and output lengths are set to $512$ and $1$, respectively. As shown in Figure~\ref{fig:ablation_spmm}, SOMM and SIMM reduce communication by $1.41\times$ to $95.3\times$ compared to GEMM, with greater savings at higher sparsity. For example, at $90\%$ sparsity, SOMM achieves an $8.1\times$ reduction. At low sparsity, SOMM and SIMM are nearly equivalent to GEMM. SOMM and SIMM also outperform SpGEMM, reducing communication by $1400\times$ to $300\times$ as sparsity increases. This is because SpGEMM repeatedly communicates the same rows and columns, whereas SOMM and SIMM communicate them only once.

\begin{figure}[h]
    \centering
    \includegraphics[width=0.4\textwidth]{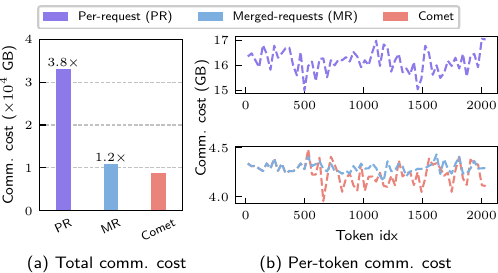}
    \caption{Communication cost comparison of cache refilling strategies over sequence lengths.}
    \label{fig:ablation_cache_fill}
    \vspace{-0.3cm}
\end{figure}

\mypara{(3) KV Cache manager.}
We compare the communication costs of three cache refilling strategies: per-request refilling (PR), merged-requests refilling (MR), and merged-requests refilling with prefetching ({\projectname}). 
To ensure consistent attention head activation for each token generation, all strategies were tested under identical input conditions. A fixed output length of 2048 was used to demonstrate long-term efficiency.
Our merging strategy significantly reduces communication costs, as illustrated in Figure \ref{fig:ablation_cache_fill}(a). MR achieves a $3.8\times$ reduction compared to PR. Additionally, the prefetching mechanism further reduces MR's communication costs by $1.2\times$.

Figure~\ref{fig:ablation_cache_fill}(b) illustrates the communication cost for generating each token. It is evident that PR incurs higher communication cost than the other two strategies. The strategy of {\projectname} is generally lower than MR but is nearly identical in the early phases (sequence length $<500$). 
This is because the sequence length is relatively short at this stage, and the cache misses length do not meet the prefetching conditions.
As the sequence extends and attention heads begin to exhibit longer misses, prefetching is triggered, reducing redundant communication.

%% file: chapters/9_related_work.tex
\section{Related Work}

\mypara{Acceleration of Private Inference for LLMs.} Current efforts to accelerate private inference for Transformers focus on two areas: model architecture and protocol optimization. For model architecture, mainstream methods replace complex non-linear functions with simpler ones to improve efficiency~\cite{chen2022x, liang2024merge, li2022mpcformer, akimoto2023privformer}. For example, THE-X~\cite{chen2022x} uses simpler ReLU functions under MPC instead of more complex functions like GeLU and Softmax, though this requires retraining to recover accuracy lost from these approximations. In protocol optimization, Iron~\cite{hao2022iron} uses a hybrid protocol with homomorphic encryption for linear layers and MPC for non-linear layers. To reduce communication overhead in non-linear layers, CipherGPT~\cite{hou2023ciphergpt}, BumbleBee~\cite{lu2023bumblebee}, and Bolt~\cite{pang2023bolt} use piecewise polynomial approximations. Since homomorphic encryption is slower than MPC, systems like Puma~\cite{dong2023puma}, SIGMA~\cite{gupta2023sigma}, SecFormer~\cite{luo2024secformer}, and SecureTLM~\cite{chen2024securetlm} rely purely on MPC and develop higher-precision approximations for non-linear functions. While these methods mainly optimize non-linear layers, {\projectname} improves both linear and non-linear functions, making it complementary to many existing approaches.

\mypara{Activation Sparsity.} Recent works leverage activation sparsity to speed up inference in plaintext LLMs. Deja Vu~\cite{liu2023deja} was the first to propose using activation sparsity to accelerate LLM inference without compromising model quality. ReLU${^2}$~\cite{zhang2024relu} and ReLU Strikes Back~\cite{mirzadeh2023relu} utilized ReLU's activation sparsity to reduce computation and weight transmission. ProSparse~\cite{song2024prosparse} further enhanced this by replacing non-ReLU activations with ReLU to exploit sparsity. PowerInfer~\cite{song2023powerinfer} preloads frequently activated neurons onto the GPU, while keeping infrequently activated ones on the CPU for acceleration. To our knowledge, {\projectname} is the first private inference system to accelerate by leveraging activation sparsity.

%% file: chapters/11_conclusion.tex
\section{Conclusion}

This paper proposes {\projectname}, an efficient private inference system leveraging activation sparsity. By predicting the sparsity of activation outputs, {\projectname} skips computations for elements predicted to be zero in both non-linear and linear layers, effectively reducing communication costs and accelerating inference. We implement {\projectname} and evaluate it across four model sizes, achieving a $1.87\times$ to $2.63\times$ speedup over state-of-the-art private inference systems.

\section*{Acknowledgements}
We sincerely thank our reviewers and shepherd for their constructive suggestions and guidance.
This work is supported by the National Natural Science Foundation of China for Distinguished Young Scholars (62125208), the National Key R\&D Program of China (2023YFB4503200), the Strategic Priority Research Program of Chinese Academy of Sciences (XDB0690100), and NSFC (62202464, 92270204 and 92267203).

%% file: chapters/7_security.tex
\section{Security Proof}
\label{sec:security_analysis}
The security of {\projectname} adheres to the standard simulation-based security definition in MPC: a protocol is considered secure if there exists a polynomial-time simulator $\mathcal{S}$ that can construct a simulated world where the view of an adversary $\mathcal{A}$ is computationally indistinguishable from the real-world view~\cite{lindell2017simulate}. 
Based on Universal Composability (UC) framework~\cite{canetti2001universally}, to prove the security of {\projectname}, it suffices to prove the security of its three core protocols: Oblivious Shuffle, SOMM, and SIMM. 

\begin{theorem}
    The protocols oblivious shuffle, SOMM and SIMM are secure against the honest-but-curious adversary. 
\end{theorem}

\mypara{Oblivous Shuffle}.
The adversary's view in the oblivious shuffle protocol is $\text{view}_\mathcal{A} = \{\rho, \tau, \ba, \bb, \shareB{x}, \tau(\shareB{x}), \shareB{y}, $  $ \rho(\shareB{y}),\shareB{z}\}$. where $\rho$ and $\tau$ are random permutations, and $\ba, \bb$ are uniformly random vectors. The secret-shared variables $\shareB{x}$ and $\shareB{y}$, as well as their shuffled results $\tau(\shareB{x})$ and $\rho(\shareB{y})$, are uniformly random due to the properties of secret sharing and random permutations. The output $\shareB{z}$, computed as the sum of two secret-shared vectors, also retains uniform randomness. A simulator $\mathcal{S}$ can replicate $\text{view}_\mathcal{A}$ by generating these components independently, ensuring that the simulated view is computationally indistinguishable from the real view. Thus, this protocol is secure against the honest-but-curious adversary.

\mypara{SOMM and SIMM}.
The adversary's view in SOMM protocol is $\text{view}_\mathcal{A} = \{\shareB{X}, \shareB{Y}, \mathcal{S}, \shareB{X}^t,  $ $\shareB{Y}^t, \shareB{R}^t, \shareB{Z}\}$. The secret-shared input $\shareB{X}, \shareB{Y}$ are uniformly random due to the properties of secret sharing. $\mathcal{S}$ is a plaintext sparsity distribution obtained by applying an oblivious shuffle to a secret-shared 0-1 sparsity distribution, where the randomness of the shuffle ensures that $\mathcal{S}$ is uniformly random. $\shareB{X}^t$ is obtained by performing local indexing operations on the secret-shared input, which do not alter their uniform randomness. $\shareB{R}^t$ is generated using a proven secure multiplication protocol, ensuring that it is also uniformly random. The output $\shareB{Z}$ is obtained by merged from secret-shared matrices, also retains uniform randomness. Thus, the SOMM protocol is secure against the honest-but-curious adversary. The security proof of SIMM follows the same reasoning as SOMM, as both protocols share a similar structure and rely on the same security guarantees.

%% file: chapters/appendix_optimality.tex
\section{Optimality Proof}

\subsection{The proof of Theorem 1}
\label{sec:appendix_somm}

We model the matrix multiplication $\mathbf{X}\mathbf{Y} = \mathbf{Z}$ using a bipartite graph $G = (X, Y, Z)$, where nodes $x \in X$ represent the row vectors of $\mathbf{X}$, nodes $y \in Y$ represent the column vectors of $\mathbf{Y}$, and edges $z \in Z$ correspond to nonzero elements in $\mathbf{Z}$, computed as the dot product of $x$ and $y$. Since zero elements in $\mathbf{Z}$ do not induce edges in $G$, some nodes in $X$ and $Y$ may be isolated, meaning they do not contribute to computation or communication. Without loss of generality, we focus on the reduced subgraph $G^r = (X^r, Y^r, Z)$, where $X^r$ and $Y^r$ include only nodes that participate in at least one multiplication.

The communication cost of an MPC matrix multiplication is given by $\mathcal{C}_1(G) = 2(|X| + |Y|)$, representing the number of row and column vectors communicated. The computation cost is $\mathcal{C}_2(G) = 3|X||Y|$, representing the total number of local dot product operations performed. For simplicity, we omit the coefficients. To achieve the minimal communication cost and, under this condition, the minimal computation cost, the problem can be formulated as finding a subgraph partition $G_1, G_2, \ldots, G_k$ of $G^r$ such that
\begin{align*}
     \min_{G_1, G_2, \ldots, G_k} & \quad \mathcal{S}_2 = \sum_{i=1}^k \mathcal{C}_2(G_i) 
    \\
    \text{s.t.} & \quad \mathcal{S}_1 = \sum_{i=1}^k \mathcal{C}_1(G_i) \text{ is minimized},
    \\
    & \quad \forall i \neq j, Z_i \cap Z_j = \varnothing.
\end{align*}

$\mathcal{S}_1$ achieves its minimum when each connected component of $G^r$ is fully assigned to a single subgraph. Since 
\begin{align*}
    \sum \mathcal{C}_1 (G_i) = \sum |X_i| + \sum |Y_i| \geq (|X^r| + |Y^r|),
\end{align*}
the minimum is exactly the total number of nodes. This minimum is achieved only if each node is assigned to a single subgraph, meaning that the connected component containing the node must be entirely within one subgraph. Otherwise, some nodes would be assigned to multiple subgraphs, increasing $\mathcal{C}_1$.

Under the condition that each connected component is assigned to a single subgraph (minimizing $\mathcal{S}_1$), the $\mathcal{S}_2$ is minimized by further ensuring that each subgraph contains exactly one connected component. Consider a subgraph $G' = (X', Y', Z')$ containing multiple connected components $G_i = (X_i, Y_i, Z_i)$ and $G_j = (X_j, Y_j, Z_j)$. The computation cost for $G'$ is:
\begin{align*}
    \mathcal{C}_2(G') 
    &= |X'||Y'| = |X_i + X_j|\cdot |Y_i + Y_j| \\ 
    & \ge |X_i||Y_i| + |X_j||Y_j|  = \mathcal{C}_2(G_i) + \mathcal{C}_2(G_j)
\end{align*}
Thus, splitting $G'$ into $G_i$ and $G_j$ reduces the computation cost. Repeating this argument, the optimal partition occurs when each connected component forms its own subgraph.

Finding all connected components of $G^r$ can be achieved using Depth-First Search (DFS). The SOMM protocol uses DFS to identify connected components, performs matrix multiplications for each component, and merges the results. Thus, Protocol~\ref{protocol:SOMM} achieves both minimal communication cost and computation optimality, completing the proof.


\subsection{The proof of Theorem 2}
\label{sec:appendix_simm}

\mypara{Communication minimization}. In Protocol~\ref{protocol:SIMM}, we adopt a column-by-row multiplication strategy. Each nonzero element in $\mathbf{X}$ is assigned to at most one column subvector, and each row of $\mathbf{Y}$ participates in at most one multiplication. Therefore, every nonzero element of $\mathbf{X}$ and every row of $\mathbf{Y}$ are masked and communicated exactly once, achieving minimal communication cost.

\mypara{Computation optimality}. While MPC inherently decomposes one matrix multiplication into multiple local multiplications, the dimensions of these local multiplications remain consistent with those of the original matrices. Therefore, the total computation cost of MPC matrix multiplication can be directly represented by the number of scalar multiplications in the original matrix multiplication. Let \(\mathbf{X} \in \mathbb{R}^{m \times n}\) and \(\mathbf{Y} \in \mathbb{R}^{n \times p}\). Define \(R_i\) as the number of nonzero elements in the \(i\)-th row of \(\mathbf{X}\) and \(C_j\) as the number of nonzero elements in the \(j\)-th column of \(\mathbf{X}\). The total number of nonzero elements is \(N = \sum_{i=1}^m R_i = \sum_{j=1}^n C_j\). Under a row-by-column approach, each nonzero element participates in \(p\) scalar multiplications, leading to \(\sum R_i \cdot p = N \cdot p\). Under a column-by-row approach, each nonzero element also contributes to \(p\) scalar multiplications, giving \(\sum C_j \cdot p = N \cdot p\). Hence, either method produces \(N \cdot p\) scalar multiplications, which matches the lower bound for computing \(\mathbf{X}\mathbf{Y}\). 

Since the column-by-row approach simultaneously achieves this computational optimum and incurs the minimal communication overhead, Protocol~\ref{protocol:SIMM} is optimal in both communication and computation, completing the proof.

%% file: chapters/appendix_generality.tex
\section{Generality of {\projectname}}
\label{sec:generality}

\mypara{Architectures.}
We evaluated {\projectname} on various Transformer architectures, including encoder-only (ViT~\cite{dosovitskiy2020image}, Bert~\cite{bert}), encoder-decoder (T5~\cite{raffel2020exploring}), and decoder-only (OPT~\cite{zhang2022opt}), all of which use ReLU. As shown in Table~\ref{tab:generality_architecture}, compared to Puma~\cite{dong2023puma}, {\projectname} achieves a 1.25$\times$-2.84$\times$ speedup in TTFT (Time To First Token).

\begin{table*}[ht]
    \centering
    \caption{Performance across different model architectures.}
    \begin{tabular}{lcccccccc}
    \toprule
    {Model} & {ViT-Base} & {ViT-Large} & {Bert-Base} & {Bert-Large} & {T5-Base} & {T5-Large} & {OPT-6.7B} & {OPT-13B} \\
    \midrule
    Sparsity (\%)  & 82.2  & 86.7  & 85.2  & 87.3  & 92.1  & 94.7  & 90.1  & 93.5  \\
    TTFT (min) & 0.85 & 2.12 & 1.34 & 2.23 & 1.97 & 3.72 & 6.5 & 10.4 \\
    Speedup  & 1.55$\times$ & 1.64$\times$ & 1.73$\times$ & 1.97$\times$ & 1.85$\times$ & 2.24$\times$ & 2.57$\times$ & 2.84$\times$ \\
    \bottomrule
    \end{tabular}
    \label{tab:generality_architecture}
\end{table*}

\begin{table*}[t]
    \centering
    \caption{Performance across different activation functions.}
    \begin{tabular}{l|ccc|ccccc}
    \toprule
    {Model} & {GPT2-XL} & {Falcon-7B} & {Bloom-7B} & {Llama2-7B~} & {Llama3-8B} & {Mixtral-7B} & {Qwen1.5-7B} & {DeepSeek-7B} \\
    \midrule
    Activation & GeLU & GeLU & GeLU & Swish & Swish & Swish & Swish & Swish \\
    Sparsity (\%) & 84.3 & 94.9 & 88.2 & 89.4 & 90.7 & 90.2 & 86.2 & 87.4 \\
    Accuracy deviation (\%) & -2.1 & -0.3 & +0.5 & -1.0 & -0.4 & +0.3 & -0.5 & -0.6 \\
    Speedup   & 1.82$\times$ & 3.12$\times$ & 2.52$\times$ & 2.69$\times$ & 2.75$\times$ & 2.31$\times$ & 2.53$\times$ & 2.62$\times$ \\
    \bottomrule
    \end{tabular}
    \label{tab:generality_activation}
\end{table*}

\mypara{Activation functions.}
We extend {\projectname} beyond ReLU-based LLMs by evaluating models that use GeLU~\cite{hendrycks2016gaussian} and SwiGLU~\cite{shazeer2020glu}, two of the most widely adopted activation functions in modern LLMs. Specifically, GeLU is used in models such as GPT2~\cite{radford2019language}, Falcon~\cite{almazrouei2023falcon}, and BLOOM~\cite{le2023bloom}, while SwiGLU is used in Llama2~\cite{touvron2023llama}, Llama3~\cite{grattafiori2024llama}, Mixtral~\cite{jiang2024mixtral}, Qwen1.5~\cite{bai2023qwen}, and DeepSeek~\cite{bi2024deepseek}.

Although these activation functions do not inherently exhibit sparsity, prior works~\cite{mirzadeh2023relu, zhang2024relu, song2024prosparse, song2024turbo, wang2024q, zhangr} have shown that replacing them with ReLU-family functions (e.g., ReLU~\cite{nair2010rectified}, ReLU$^2$\cite{zhang2024relu}, shiftedReLU\cite{mirzadeh2023relu}, dReLU~\cite{song2024turbo}) followed by fine-tuning enables activation sparsity without significant accuracy loss. We applied this ReLUfication strategy to eight different LLMs, as shown in Table~\ref{tab:generality_activation}. Each model achieved an 80\%-94\% sparsity ratio, with an average accuracy drop of less than 0.6\% on the GLUE benchmark. On these sparsified LLMs, {\projectname} achieves a 1.82×-3.12× speedup over Puma.

%% file: chapters/appendix_baselines.tex
\section{MPC-based Private Inference Systems}
\label{sec:appendix_baselinse}

\mypara{MPCFormer} accelerates Transformer private inference by replacing costly Softmax and GeLU functions with simple quadratic approximations. Since these approximations degrade model accuracy, knowledge distillation is applied to restore performance. While effective in reducing MPC overhead, this approach introduces additional training steps.

\mypara{SecFormer}, like MPCFormer, replaces Softmax with a quadratic approximation but further improves efficiency by designing a high-precision polynomial for GeLU and an MPC-friendly LayerNorm computation. To compensate for approximation errors, it also requires fine-tuning the model.

\mypara{Puma} enhances MPC efficiency by designing high-precision polynomial approximations for Softmax and GeLU, ensuring a slight accuracy loss. It also introduces secure protocols for Embedding and LayerNorm computations. Unlike other approaches, it achieves near plaintext-level accuracy in encrypted inference without requiring any model fine-tuning.

The above works adopt semi-honest security. Some, like MUSE~\cite{lehmkuhl2021muse} and SIMC~\cite{chandran2022simc}, achieve malicious security by ensuring protocol termination upon an attack, preventing sensitive information leakage. However, this significantly reduces inference efficiency. Enhancing malicious security in private LLM inference remains a key direction for future research.

%% file: chapters/10_discussion.tex
\section{Impact of Revealing Sparsity Levels}
\label{appendix:dp}
{\projectname} leverages activation sparsity to accelerate private inference. Compared to existing private inference systems, the only additional information it reveals is the activation sparsity levels. Although there are currently no known effective attacks targeting sparsity levels, this information could introduce potential privacy risks. For example, in a semi-honest setting, an adversary could issue repeated inference requests for different inputs to map the relationship between input and sparsity level. In a malicious setting, a model provider could design a predictor that exhibits specific sparsity levels on some layers for certain inputs. Notably, the risks discussed here are not unique to {\projectname}; these inference attacks~\cite{carlini2022membership} and backdoor attacks~\cite{li2022backdoor} remain open challenges even for state-of-the-art private inference systems.

Fortunately, differential privacy (DP)~\cite{NIPS2015_a0161022} under MPC can effectively mitigate risks of revealing sparsity levels. Specifically, before revealing  the shuffled sparsity distribution  $\shareB{s}$, the MPC servers collaboratively generate a secret-shared 0-1 perturbation vector $\shareB{p}$, where the number of ones is determined via MPC-based DP. Then, by applying an XOR operation between  $\shareB{s} $ and  $\shareB{p} $, some ``0" are flipped to ``1", hidding the exact sparsity level. Importantly, this transformation does not turn ``1" into ``0", ensuring that all activated neurons are still computed, preserving model accuracy. We evaluate the performance of {\projectname} with DP. For Llama2-7B, even under strong DP guarantees ($\epsilon$  = 0.01), {\projectname} achieves 71.1\% sparsity and a 2.01× speedup over Puma. 

\begin{table}[h]
    \centering
    \caption{The performance of {\projectname} with DP under different privacy budget.}
    \label{tab:dp}
    \begin{tabular}{lccccc}
        \toprule
        \textbf{Privacy budget} ($\epsilon$)  & w/o DP  & 0.5  & 0.1  & 0.05  & 0.01  \\
        \midrule
        \textbf{Sparsity} (\%)  & 89.4 & 87.7 & 85.9 & 79.9 & 71.7 \\
        \textbf{Speedup}        & 2.69$\times$ & 2.63$\times$ & 2.52$\times$ & 2.34$\times$ & 2.01$\times$ \\
        \bottomrule
    \end{tabular}
\end{table}

%% file: chapters/meta_review.tex
\newpage 

\section{Meta-Review}

The following meta-review was prepared by the program committee for the 2025
IEEE Symposium on Security and Privacy (S\&P) as part of the review process as
detailed in the call for papers.

\subsection{Summary}
The authors propose {\projectname}, an MPC-based private inference system for LLMs that exploits activation sparsity to reduce communication and computation overhead. The authors introduce new protocols to securely skip zero-valued neurons and develop a KV-cache management strategy compatible with sparse inference. Extensive experiments demonstrate significantly reduced inference time (1.87x-2.63x) and communication (1.94x-2.64x) compared to state-of-the-art private LLM inference systems, with further evidence showing generalization across different activation functions, architectures, and tasks.

\subsection{Scientific Contributions}
\begin{itemize}
\item Provides a Valuable Step Forward in an Established Field
\item Creates a new tool to enable Future Science
\end{itemize}

\subsection{Reasons for Acceptance}
\begin{enumerate}
\item The paper addresses a timely challenge in secure inference for LLMs and notably improves over state-of-the-art in speed and communication.
\item The proposed method is presented comprehensively, and the authors provide an extensive evaluation (including WAN settings, different activation functions, and various LLM architectures).
\item The paper contains many interesting and novel ideas on how to speed up inference.
\end{enumerate}